\documentclass[aps,prd,10pt,twocolumn,superscriptaddress,nofootinbib]{revtex4-1}

\usepackage{amsmath,amssymb,bm}
\usepackage{graphicx}
\usepackage{booktabs}
\usepackage{dcolumn}
\usepackage{microtype}
\usepackage[colorlinks=true,linkcolor=blue,citecolor=blue,urlcolor=blue]{hyperref}

\newcommand{\dd}{\mathrm{d}}
\newcommand{\DM}{\mathrm{DM}}
\newcommand{\rhDM}{\rho_{\DM}}

\begin{document}

\title{Scalar dark matter in space-based gravitational-wave detectors: center-of-mass motion, size breathing, and TDI projection}

\author{Rui-Yang Hu}
\author{Yuan-Zhi Li}
\author{An-Qi Wang}
\author{Zong-Ru Zou}
\author{Fa-Peng Huang}
\author{Cheng-Gang Qin}
\email{qinchg3@sysu.edu.cn}
\affiliation{MOE Key Laboratory of TianQin Mission, TianQin Research Center for Gravitational Physics \& School of Physics and Astronomy, Frontiers Science Center for TianQin, Gravitational Wave Research Center of CNSA, Sun Yat-sen University (Zhuhai Campus), Zhuhai 519082, China}

\date{\today}

\begin{abstract}
Ultralight scalar dark matter can make space-based gravitational-wave detectors respond through both the scalar charge of freely falling test masses and scalar-induced changes of local solid length scales.  Existing space-detector forecasts usually model the former as a center-of-mass force, while ground-based interferometer studies show that scalar fields can also act through material and optical-path transduction.  We ask which part of a local material response survives after one-way Doppler measurements are assembled into delayed time-delay-interferometry observables.  To this end, we formulate center-of-mass motion and endpoint-size breathing in a common link-response notation for LISA-, Taiji-, and TianQin-like detectors.  The main result is a projection rule: in the equal-arm, identical-endpoint, common-field limit, endpoint breathing enters Michelson-$X$ as a common-mode link perturbation and is removed from the retained channel.  Its leading leakage is controlled by finite scalar wave vector, unequal or time-dependent arms, nonidentical endpoint response, or auxiliary readouts, and carries extra geometric and delay suppressions beyond the local size response.  We then give reproducible noise, sensitivity, and network-combination formulas, and quote multi-mission improvements only under explicit independent-stream and scalar-coherence assumptions.  The result provides a controlled baseline for deciding when test-mass breathing can be neglected and when instrument-specific material response must be modeled.
\end{abstract}

\maketitle

\section{Introduction}

The microscopic nature of dark matter remains unknown.  A broad and well-motivated class of candidates consists of ultralight bosonic fields whose large occupation number allows them to behave as coherent classical waves over macroscopic coherence lengths and times \cite{Marsh2016,Hui2017Fuzzy,Graham2016DMReview,MillerReview2025}.  Scalar or dilaton-like versions of this idea are especially relevant for precision measurements because they can couple to Standard Model parameters and therefore act as oscillating variations of fundamental constants \cite{DamourDonoghue2010,Arvanitaki2015DM,StadnikFlambaum2015,Alachkar2025PRL}.  Such variations have been searched for, or constrained by, atomic clocks, atomic spectroscopy, resonant cavities, equivalence-principle tests, and laser interferometers \cite{DereviankoPospelov2014,VanTilburg2015,Hees2016,Hees2018arXiv,Savalle2021,GroteStadnik2019,Vermeulen2021,Aiello2022,Goettel2024PRL}.

Space-based gravitational-wave interferometers provide a natural low-frequency laboratory for this class of signals.  LISA, Taiji, and TianQin use freely falling test masses and million- to hundred-thousand-kilometer baselines to measure optical phase variations in the millihertz band \cite{LISA2017,Taiji2017,TianQin2016}.  Their target band corresponds to scalar masses around $m_\phi \sim 10^{-19}$--$10^{-15}\,\mathrm{eV}$, where the dark-matter field is a narrow-band, highly occupied classical wave.  The detector modeling used for this problem builds on the standard space-interferometer response and TDI literature, including the cancellation of laser noise, the construction of synthetic Michelson combinations, and equal-arm sensitivity estimates \cite{TintoArmstrong1999,EstabrookTintoArmstrong2000,LarsonHiscockHellings2000,LarsonHellingsHiscock2002,CornishRubbo2003,Prince2002,TintoDhurandhar2014,RobsonCornishLiu2019,Ren2023TDC}.  Recent calculations have developed the response of space laser interferometers to ultralight scalar, vector, gravitational, stochastic, and axionlike dark-matter channels \cite{Yu2023PRD,YuEtAl2024Gravitational,YaoTang2024Stochastic,JiangTang2026ULDMInterferometers,LiuEtAl2026EPJC}; related work has examined how the multi-channel structure, angular response, and orbital modulation of LISA- and Taiji-like detectors can help distinguish ultralight-field signals from quasi-monochromatic gravitational waves \cite{Cutler1998,Xu2025Monochromatic,GueWolfHees2025LISA,YaoEtAl2025Spectral}.

There is also a broader gravitational-wave-detector dark-matter literature that is complementary to the scalar response considered here.  Ground-based and proposed interferometers have been studied as transducers for axionlike fields, ultralight vector dark matter, light dark-matter scattering, and scalar-induced material strain in auxiliary cavities or optical components \cite{Nagano2019PRL,Nagano2021PRD,Michimura2020,LeeNugrohoSpinrath2020EPJC,AggarwalHall2022}.  Other work uses gravitational-wave observations to probe dark-sector cosmology, compact-object environments, or axion-motivated stochastic backgrounds rather than direct material couplings to the detector \cite{Bandyopadhyay2025AxionGW,ChenWangLuoShao2026Review,MillerReview2025}.  The present paper belongs to the direct detector-response line: it asks how a scalar-induced material response enters the one-way links and the delayed data combinations used for laser-noise suppression.

The motivation for the present work comes from the different way in which ground- and space-based interferometers turn material motion into data.  In ground-based Michelson interferometers, scalar-induced motion of mirrors or other optical elements, together with optical-index changes, can be converted into differential optical length signals when the readout weights the two arms or the relevant material surfaces asymmetrically \cite{StadnikFlambaum2015,GroteStadnik2019,Vermeulen2021,Aiello2022,Fukusumi2023PRD,Goettel2024PRL,LVK2025MultiModel}.  Here ``asymmetric'' means that nominally common material responses do not enter the final readout with identical delayed weights, signs, and optical transfer functions.  Space missions record one-way Doppler links between separated spacecraft, and TDI data combinations are then synthesized by applying time delays and linear combinations to those links.  A local surface displacement that looks sizeable in a single-link estimate can therefore be projected into a common-mode part of a TDI combination, while a smaller local effect may survive if it carries the phase and delay structure selected by that combination.

The question addressed here is therefore not only whether scalar dark matter moves a test-mass surface, but whether that motion survives the delayed-link combinations used by LISA, Taiji, and TianQin.  We separate two scalar-DM responses that are often discussed together.  The first is the center-of-mass (CM) motion of test masses caused by the gradient of the scalar field.  The second, modeled explicitly below, is the physical breathing of the local test-mass endpoint caused by scalar-induced changes of atomic length scales.  We show that the endpoint-size channel is a common-mode-dominated contribution in the equal-arm common-field limit, and we identify the conditions under which it can become measurable: finite dark-matter wave vector, unequal or time-dependent arms, nonidentical material response, or dedicated auxiliary readouts.  In the standard Au-Pt test-mass model, the endpoint-size residual is not generically the leading contribution for LISA, Taiji, or TianQin.

Our analysis is complementary to previous interferometer searches and forecasts, but it asks a slightly different question.  The work of Stadnik and Flambaum, and later Grote and Stadnik, developed the material-response picture for laser interferometers and showed that scalar dark matter can enter through mirror motion, beam-splitter motion, optical-index changes, and laser-frequency effects \cite{StadnikFlambaum2015,GroteStadnik2019}.  GEO600, the Fermilab Holometer, LIGO, and related transducer proposals then use real interferometers, cavities, or auxiliary optical paths to search for these effects, with the final observable determined by the optical layout and by small but important asymmetries between nominally similar components \cite{Savalle2021,Vermeulen2021,Aiello2022,AggarwalHall2022,Fukusumi2023PRD,Goettel2024PRL,LVK2025MultiModel}.  In those experiments it is natural to emphasize how scalar-induced size or index changes become differential length signals.

Space-based detectors have a different readout structure.  LISA, Taiji, and TianQin measure one-way Doppler frequency shifts rather than the instantaneous difference between two local arms.  TDI data combinations are subsequently synthesized from delayed link data so that the otherwise dominant laser-frequency noise is rejected \cite{TintoArmstrong1999,TintoDhurandhar2014}.  Yu et al. used one-way Doppler observables and transfer functions of TDI combinations to calculate the sensitivity of space interferometers to ultralight scalar, vector, and gravitational-wave dark matter \cite{Yu2023PRD}; related work has since examined gravitational, stochastic, and axionlike ultralight-field channels \cite{YuEtAl2024Gravitational,YaoTang2024Stochastic,LiuEtAl2026EPJC}, instrument-level laser, clock, and optical-bench response channels \cite{JiangTang2026ULDMInterferometers}, and channel-based discrimination in realistic space-detector data streams \cite{Xu2025Monochromatic,GueWolfHees2025LISA,YaoEtAl2025Spectral}.  The present paper inserts the material-size response into the same sequence of link response and delayed data combination.  It then asks whether this response lies in the subspace retained by the chosen TDI combination or in the common-mode null space.

This distinction matters for interpretation.  A size change that would be observable in a ground-based asymmetric optical layout can be removed in a symmetric delayed-link combination.  Conversely, a contribution that is suppressed at the level of a local endpoint may survive if it carries the phase and delay structure selected by that combination.  Therefore the relevant comparison is not just between two local displacement scales, $R K_R\phi_0$ and $\alpha_A k\phi_0/\omega_\phi^2$, but between their projected transfer functions in the actual data channel.

Table~\ref{tab:related} makes this positioning explicit and fixes the scope of the paper.  The first row records the material-response physics established by ground-based interferometer work: scalar fields can move optical elements and modify optical paths.  The second row identifies the space-detector ingredients that we adopt: one-way Doppler links, transfer functions of TDI combinations, and equal-arm noise normalizations.  The third row explains why channel content and modulation matter for interpreting monochromatic signals.  The last row states the specific contribution of this work, namely the decomposition of the scalar response into CM motion and endpoint-size breathing, followed by projection through the Michelson-$X$ combination.  The table is therefore a roadmap for the calculation rather than a sensitivity ranking.

\begin{table*}[t]
\caption{Representative nearby work and the role it plays in the present analysis.  The table is not a ranking of sensitivities; it separates the physical questions that are often mixed together when scalar-DM interferometer responses are compared.}
\label{tab:related}
\begin{ruledtabular}
\begin{tabular}{lll}
Work class & Main observable & Relation to this paper\\
\hline
\begin{minipage}[t]{0.22\textwidth}\raggedright Material-response theory and ground-based transducers \cite{GroteStadnik2019,Vermeulen2021,Aiello2022,Goettel2024PRL,LVK2025MultiModel}\end{minipage}
& \begin{minipage}[t]{0.24\textwidth}\raggedright Local optical length, mirror or beam-splitter motion, and index changes in asymmetric readouts\end{minipage}
& \begin{minipage}[t]{0.43\textwidth}\raggedright Establishes that scalar DM can make solids and optical paths oscillate.  We ask whether the corresponding size response survives the delayed common-mode projection of a space-detector data combination.\end{minipage}\\[1.0ex]
\begin{minipage}[t]{0.22\textwidth}\raggedright Space-interferometer ultralight-field forecasts \cite{Yu2023PRD,YuEtAl2024Gravitational,YaoTang2024Stochastic,LiuEtAl2026EPJC}\end{minipage}
& \begin{minipage}[t]{0.24\textwidth}\raggedright One-way Doppler links and TDI-combination responses of scalar, vector, gravitational, or stochastic fields\end{minipage}
& \begin{minipage}[t]{0.43\textwidth}\raggedright Provides the natural noise normalization and transfer functions.  We decompose the scalar response into CM motion and material-size breathing before applying the same delayed combinations.\end{minipage}\\[1.0ex]
\begin{minipage}[t]{0.22\textwidth}\raggedright Space-detector signal discrimination studies \cite{Xu2025Monochromatic,GueWolfHees2025LISA,YaoEtAl2025Spectral}\end{minipage}
& \begin{minipage}[t]{0.24\textwidth}\raggedright Channel structure, orbital modulation, and spectral differences between ULDM and monochromatic gravitational waves\end{minipage}
& \begin{minipage}[t]{0.43\textwidth}\raggedright Shows that the channel content of a signal is phenomenologically important.  Our common-mode projection of the size response supplies a material-response example of such channel dependence.\end{minipage}\\[1.0ex]
\begin{minipage}[t]{0.22\textwidth}\raggedright This work\end{minipage}
& \begin{minipage}[t]{0.24\textwidth}\raggedright CM response plus the first nonzero residual of scalar-induced size breathing after Michelson-$X$ projection\end{minipage}
& \begin{minipage}[t]{0.43\textwidth}\raggedright Identifies when endpoint breathing is projected into the common-mode part of Michelson-$X$ and quantifies the leading residual within a baseline Au-Pt endpoint model.\end{minipage}\\
\end{tabular}
\end{ruledtabular}
\end{table*}

This map also clarifies what is deliberately not included.  We use the ground-based literature to define local material responses, but we do not import a ground-based optical-layout transfer function into a space detector.  We use the space-interferometer literature to define the one-way link response, TDI combinations, and noise normalization, but we add a separate endpoint-size response before passing it through Michelson-$X$.  This is why the quantitative forecast below is restricted to the test-mass endpoint model and to the CM response.

The paper is organized as follows.  Section II defines the scalar field and the material coefficients used for CM and size responses.  Section III compares the two effects at the single-link level, where the size channel can look deceptively large.  Section IV shows how a symmetric size response is projected by Michelson-$X$ and identifies the first common-mode-breaking terms.  Section V gives equal-arm baseline sensitivities and network forecasts for LISA, Taiji, and TianQin.  Section VI discusses what would be needed to turn the size channel into a leading search channel.  Section VII states the assumptions that must be relaxed in a final mission-level forecast, Sec. VIII concludes, and the Appendices give the delay-operator algebra and the external-limit conversion used in Fig.~\ref{fig:external}.

\section{Scalar field and material response}

\subsection{Classical scalar dark matter}

Throughout the paper we use natural units, $\hbar=c=1$.  Mass, angular frequency, inverse time, and inverse length are therefore expressed in the same units.  The scalar mass $m_\phi$ is quoted in eV in figures and tables.  Engineering parameters such as arm lengths and instrumental noise amplitudes are listed in conventional units for traceability, but they are converted to natural units before being inserted into the response and noise formulas.

We model the local scalar field as a nonrelativistic coherent wave,
\begin{equation}
  \phi(t,\bm{x}) = \phi_0 \cos(\omega_\phi t-\bm{k}\cdot\bm{x}+\varphi).
  \label{eq:field}
\end{equation}
Here $\phi_0$ is the field amplitude, $\omega_\phi$ is the scalar angular frequency, $\bm{k}$ is the dark-matter wave vector, and $\varphi$ is an arbitrary phase.  For the nonrelativistic dark-matter wave,
\begin{equation}
  \omega_\phi\simeq m_\phi,\qquad
  |\bm{k}|=m_\phi v,
  \label{eq:dispersion}
\end{equation}
where $v\sim10^{-3}$ is the characteristic dimensionless Galactic virial velocity.  If the scalar constitutes the local dark-matter density $\rho_{\rm DM}$, the canonical field amplitude is
\begin{equation}
  \phi_0 = \frac{\sqrt{2\rhDM}}{m_\phi}.
  \label{eq:phi0}
\end{equation}
All observable amplitudes below depend on dimensionless combinations such as $\phi_0/\Lambda_i$ or $\alpha_A\phi_0$.  In the response formulas and in the formulas for TDI combinations, $\omega$ denotes the Fourier angular frequency and is set to $\omega_\phi$ for a monochromatic scalar-DM signal.  The field coherence time is taken to be
\begin{equation}
  \tau_{\rm coh}\simeq \frac{2\pi}{\omega_\phi v^2}
  =\frac{1}{f_\phi v^2},
  \label{eq:cohtime}
\end{equation}
where $f_\phi\equiv\omega_\phi/(2\pi)$ is the cyclic scalar frequency.  In the analytic formulas it is a natural-unit quantity; values in Hz are obtained only at the plotting or reporting stage by restoring $\hbar$.  This coherence time becomes important when the observing time is comparable to or longer than $\tau_{\rm coh}$.

\subsection{Linear couplings}

We use the standard linear scalar coupling parameterization \cite{GroteStadnik2019,DamourDonoghue2010}.  For the electron, photon, and one-parameter nucleon sectors used in the forecast, the relevant interaction terms may be written as
\begin{equation}
  {\cal L}_{\rm int}
  \supset
  -\frac{\phi}{4\Lambda_\gamma}F_{\mu\nu}F^{\mu\nu}
  -\frac{\phi}{\Lambda_e}m_e\bar e e
  -\frac{\phi}{\Lambda_N}m_N\bar N N .
  \label{eq:lint}
\end{equation}
Here $F_{\mu\nu}$ is the electromagnetic field tensor, $e$ and $N$ denote the electron and nucleon fields, and $\Lambda_\gamma$, $\Lambda_e$, and $\Lambda_N$ are the microscopic scalar-coupling scales for the photon, electron, and nucleon sectors.  To first order in the weak scalar field, these terms are equivalent to promoting the constants in the electromagnetic and mass terms to slowly varying local quantities,
\begin{equation}
\begin{aligned}
  \alpha(\phi)&=\alpha\left(1+\frac{\phi}{\Lambda_\gamma}\right),\\
  m_e(\phi)&=m_e\left(1+\frac{\phi}{\Lambda_e}\right),\\
  m_N(\phi)&=m_N\left(1+\frac{\phi}{\Lambda_N}\right).
\end{aligned}
  \label{eq:effectiveconstants}
\end{equation}
The fractional variations used below therefore are
\begin{equation}
  \frac{\delta\alpha}{\alpha}=\frac{\phi}{\Lambda_\gamma},
  \qquad
  \frac{\delta m_e}{m_e}=\frac{\phi}{\Lambda_e},
  \qquad
  \frac{\delta m_N}{m_N}=\frac{\phi}{\Lambda_N}.
  \label{eq:couplings}
\end{equation}
For a macroscopic body or test mass $A$ we define
\begin{equation}
  \frac{\delta M_A}{M_A}\equiv \alpha_A\phi,
  \qquad
  \frac{\delta R_A}{R_A}\equiv K_{R,A}\phi ,
  \label{eq:macrocoeff}
\end{equation}
where $\alpha_A$ is the effective scalar charge of the material and $K_{R,A}$ is the size-response coefficient.  Both coefficients have the inverse units of the scalar field in Eq.~\eqref{eq:field}; under a single microscopic coupling they are proportional to $1/\Lambda_i$.
For the one-parameter limits shown later, it is useful to separate the microscopic inverse scale from the dimensionless material coefficient:
\begin{equation}
  \alpha_A=\sum_{i=\gamma,e,N}\frac{C_{M,i}^A}{\Lambda_i},
  \qquad
  K_{R,A}=\sum_{i=\gamma,e,N}\frac{C_{R,i}^A}{\Lambda_i}.
  \label{eq:cmcr}
\end{equation}
Here $C_{M,i}^A$ specifies how the rest energy of the whole body responds to coupling sector $i$, while $C_{R,i}^A$ specifies how the endpoint length scale responds.  In a single-coupling scan only one $1/\Lambda_i$ is nonzero, so the response power is proportional to the square of the corresponding coefficient.  This is the convention used in the numerical code and in Eq.~\eqref{eq:sens}.

The coefficient $\alpha_A$ follows from the logarithmic response of the total rest energy of the body to the constants varied in Eq.~\eqref{eq:couplings}.  Writing the material mass schematically as nucleon rest mass plus electronic and binding contributions, one has
\begin{equation}
  \alpha_A =
  \frac{Q_\gamma^A}{\Lambda_\gamma}
  +\frac{Q_e^A}{\Lambda_e}
  +\frac{Q_N^A}{\Lambda_N},
  \qquad
  Q_i^A\equiv
  \frac{\partial\ln M_A}{\partial\ln X_i},
  \label{eq:materialcharge}
\end{equation}
with $X_i=\{\alpha,m_e,m_N\}$.  For a heavy Au-Pt test mass, $Q_N^A\simeq1$ because the mass is dominated by nucleon rest energy; the electron rest-mass fraction gives $Q_e^A$ of order $10^{-4}$; and electromagnetic/nuclear binding terms give $Q_\gamma^A$ of order $10^{-3}$.  Keeping only the representative magnitudes used for the baseline interferometer material model gives \cite{GroteStadnik2019,DamourDonoghue2010}
\begin{equation}
  \alpha_A \simeq
  \frac{4\times 10^{-3}}{\Lambda_\gamma}
  +\frac{2\times 10^{-4}}{\Lambda_e}
  +\frac{1}{\Lambda_N}.
  \label{eq:auptcharge}
\end{equation}
Thus the baseline coefficients used below are $C_{M,\gamma}^A=4\times10^{-3}$, $C_{M,e}^A=2\times10^{-4}$, and $C_{M,N}^A=1$ for an Au-Pt-like test mass.  An important consequence is that, even if the nucleon coupling is absent, electron and electromagnetic binding contributions leave a nonzero scalar charge.  Therefore a vanishing CM channel is an additional model assumption, not a generic consequence of electron or photon coupling.

It is useful to keep the roles of the two coefficients conceptually separate.  The coefficient $\alpha_A$ determines how the rest mass of the whole test mass responds to the scalar field.  A spatial gradient of the scalar field then exerts a force on the test mass and moves its center of mass.  By contrast, $K_{R,A}$ describes how the internal equilibrium length scale of the solid responds to a local change in constants.  This response can move an optical surface even if the center of mass is not accelerated.  The two effects therefore have different symmetry properties: a CM force is sensitive to the relative scalar phase between separated spacecraft, while a perfectly synchronized size breathing of identical endpoints is common mode.

For the size response we use the adiabatic Bohr-radius scaling,
\begin{equation}
  \frac{\delta R}{R}\simeq
  -\frac{\delta\alpha}{\alpha}
  -\frac{\delta m_e}{m_e},
  \label{eq:sizebohr}
\end{equation}
with possible optical-index terms treated analogously when relevant \cite{GroteStadnik2019,Vermeulen2021,Aiello2022}.

Eq.~\eqref{eq:sizebohr} follows from the leading dependence of an atomic length scale on the Bohr radius, $a_0\propto(\alpha m_e)^{-1}$.  Under the single-coupling convention of Eq.~\eqref{eq:cmcr}, the corresponding endpoint coefficients are $C_{R,\gamma}^A\simeq-1$, $C_{R,e}^A\simeq-1$, and a nucleon coefficient suppressed by finite-nuclear-mass effects, $|C_{R,N}^A|\sim m_e/m_N\simeq5.4\times10^{-4}$.  The signs depend on the convention used for the endpoint normal and are immaterial for the power sensitivities, where $|C_{R,i}^A|^2$ enters.

The Bohr-radius estimate should be read as a low-frequency material coefficient rather than as a complete elastic model.  It is valid when the scalar oscillation is adiabatic compared with internal mechanical and electronic response frequencies.  At frequencies near an internal mechanical resonance, or for a detailed spacecraft optical bench, the scalar-induced strain must be passed through the appropriate elastic and optical transfer functions.  We do not include a separate beam-splitter or optical-bench transduction term in the quantitative forecast.  Unlike a ground-based Michelson interferometer, the space missions considered here use mission-specific optical benches, telescopes, phasemeters, and one-way interspacecraft links rather than a single fixed beam splitter that directly defines the science observable.  The final optical layouts and calibration conventions are sufficiently instrument dependent that such material responses should be modeled in an end-to-end mission design.  The mass range considered below corresponds to millihertz frequencies, far below the internal resonances of compact test masses, so the adiabatic endpoint coefficient is adequate for isolating the TDI symmetry question addressed in this paper.

\section{One-arm response}

\begin{figure}[!htbp]
\centering
\includegraphics[width=0.82\columnwidth]{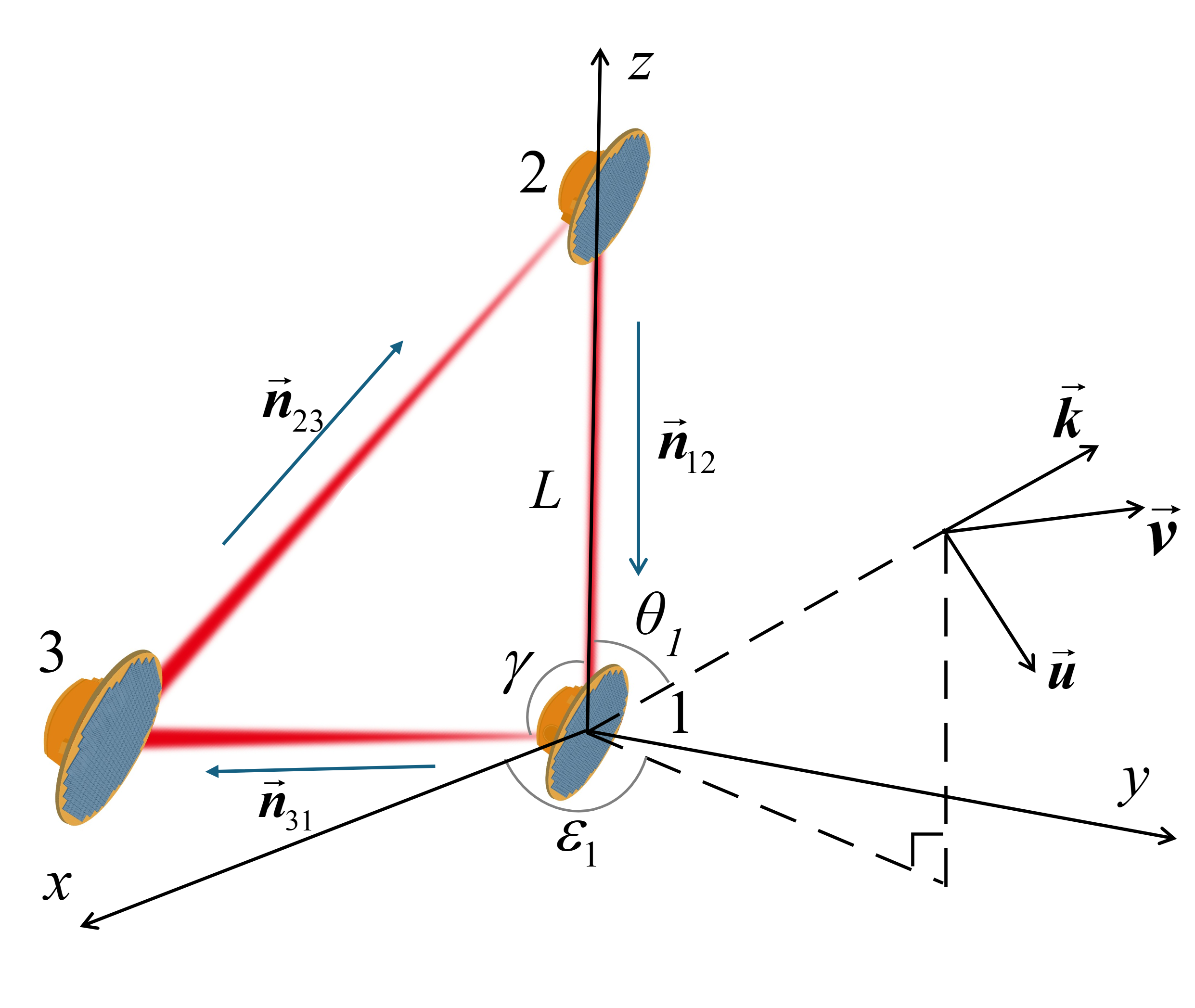}
\caption{Equal-arm triangular constellation and angular conventions used in the response formulas.  Spacecraft 1 is the Michelson-$X$ vertex, the unit vectors $\hat n_{ij}$ define the directed arms, and $\hat k$ is the scalar-field propagation direction.}
\label{fig:geometry}
\end{figure}

Figure~\ref{fig:geometry} fixes the notation used in the response calculation.  Spacecraft 1 is the vertex of the Michelson-$X$ observable in Eq.~\eqref{eq:tdix}; the directed arm vectors $\hat n_{ij}$ set the signs and link projections in the one-way Doppler response; and the scalar propagation direction $\hat k$ is the angular variable averaged over in the transfer functions below.

\subsection{Center-of-mass motion}

Let $M_A(\phi)=M_{A0}[1+\alpha_A\phi(t,\bm{x})]$, where $M_{A0}$ is the unperturbed test-mass rest energy.  In natural units, mass and rest energy have the same units.  The nonrelativistic action of a freely falling test mass may be expanded as
\begin{align}
  S_A
  =&-\int M_A(\phi)\,\dd s
  \nonumber\\
  \simeq&
  \int \dd t
  \left[
  \frac{1}{2}M_{A0}\dot{\bm X}_A^2
  -M_{A0}\left(1+\alpha_A\phi(t,\bm X_A)\right)
  \right],
  \label{eq:pointaction}
\end{align}
where terms of order $\dot{\bm X}_A^2\alpha_A\phi$ have been neglected.  Varying Eq.~\eqref{eq:pointaction} and evaluating the field gradient at the unperturbed position $\bm{x}_A$ gives the CM equation of motion
\begin{equation}
  \bm{a}_A(t) =
  -\alpha_A \bm{\nabla}\phi(t,\bm{x}_A).
  \label{eq:accel}
\end{equation}
No additional dimensional conversion factor appears because $\hbar=c=1$ has been fixed at the beginning of the paper; mass and rest energy are represented by the same quantity in Eq.~\eqref{eq:pointaction}.
For the plane wave in Eq.~\eqref{eq:field}, $\bm{\nabla}\phi=\bm{k}\phi_0\sin(\omega_\phi t-\bm{k}\cdot\bm{x}+\varphi)$.  Solving the forced equation of motion after dropping the homogeneous transient gives
\begin{equation}
  \delta\bm{X}_A^{\rm CM}(t)
  =\frac{\alpha_A\bm{k}\phi_0}{\omega_\phi^2}
  \sin(\omega_\phi t-\bm{k}\cdot\bm{x}_A+\varphi).
  \label{eq:cmx}
\end{equation}
The single-test-mass displacement scale is therefore $|\delta X_A^{\rm CM}|\sim|\alpha_A\phi_0|\,k/\omega_\phi^2=|\alpha_A\phi_0|\,v/\omega_\phi$.

For an arm whose unperturbed endpoints are spacecraft $s$ and $r$, define the directed unit vector
\begin{equation}
  \hat{\bm n}_{rs}\equiv\frac{\bm{x}_r-\bm{x}_s}{L},
  \qquad
  L=|\bm{x}_r-\bm{x}_s|.
\end{equation}
The instantaneous geometric length is
\begin{equation}
  L_{rs}(t)=
  \left|
  \bm{x}_r+\delta\bm X_r(t)
  -\bm{x}_s-\delta\bm X_s(t)
  \right|,
\end{equation}
so, to first order in the small displacements,
\begin{equation}
  \delta L_{rs}^{\rm CM}(t)
  =
  \hat{\bm n}_{rs}\cdot
  \left[
  \delta\bm{X}^{\rm CM}_r(t)
  -
  \delta\bm{X}^{\rm CM}_s(t)
  \right].
  \label{eq:deltalcm}
\end{equation}
For identical test masses the coefficient is the common material charge $\alpha_A$, not a material difference.  Material differences can add an equivalence-principle-violating term proportional to $\Delta\alpha=\alpha_s-\alpha_r$, but this is not the baseline Au-Pt configuration.

Eq.~\eqref{eq:deltalcm} also explains the $v^2$ scaling.  The scalar gradient in Eq.~\eqref{eq:accel} supplies one power $k/\omega_\phi=v$.  In the long-wavelength limit, the two endpoint displacements are nearly common and their difference supplies a second small factor $kL$; after division by $L$, the equivalent strain scale is
\begin{equation}
  h_{\rm CM}^{\rm LW}=|\alpha_A| v^2\phi_0 .
  \label{eq:hcm}
\end{equation}

\subsection{Size breathing}

The optical link endpoint is on the surface of a test mass or optical assembly rather than at its mathematical center.  Let $R_A$ denote the relevant linear size projected along a given link, and let
\begin{equation}
  R_A(t)=R_A+\delta R_A(t),
  \qquad
  \delta R_A(t)=R_A K_{R,A}\phi(t,\bm{x}_A).
  \label{eq:deltar}
\end{equation}
For a surface used by the link $s\rightarrow r$, write the scalar-induced displacement of that surface along the optical axis as
\begin{equation}
  q_A^{(rs)}(t)=\chi_A^{(rs)}\,R_A K_{R,A}\phi(t,\bm{x}_A),
  \label{eq:qendpoint}
\end{equation}
where $\chi_A^{(rs)}$ is a sign and projection factor fixed by the local optical geometry.  For the baseline endpoint estimate we take identical test masses, identical response coefficients, and $|\chi_A^{(rs)}|=1$; signs are absorbed into the link convention because the sensitivity depends on response power.  The single-link timing perturbation due only to endpoint breathing has the same delayed-endpoint structure as the CM term,
\begin{equation}
  \delta t_{rs}^{\rm size}(t)
  =
  q_r^{(rs)}(t)-q_s^{(rs)}(t-L).
  \label{eq:linksize}
\end{equation}
Thus the local one-endpoint path-length scale is
\begin{equation}
  \delta L_{\rm size}^{\rm end}(t) = K_R R\,\phi(t),
  \qquad
  h_{\rm size}^{\rm end}= \frac{R}{L}K_R\phi_0 .
  \label{eq:hsize}
\end{equation}
Here $R$ is the representative endpoint size projected along the optical link and $K_R$ is the corresponding material-response coefficient for identical endpoints.  The superscript ``end'' denotes a local endpoint path-length estimate rather than the final response after a delayed data combination, and $h_{\rm size}^{\rm end}$ is the associated path-length amplitude divided by $L$.  At this single-link level the size channel may appear competitive because $h_{\rm CM}^{\rm LW}$ carries the velocity suppression $v^2$.  This comparison is incomplete for TDI data combinations: the same endpoint perturbation can enter the delayed links as a nearly common-mode contribution and is then projected mainly into the null direction of the Michelson-$X$ combination.

This is the key distinction between a local transduction estimate and a space-detector data observable.  A single-link estimate answers the question of whether the scalar field moves a surface.  A TDI data combination asks the more restrictive question of whether that surface motion has the delayed antisymmetry needed to survive the laser-noise-canceling combination.  The answer depends on the symmetry of the constellation and of the material response.  In the idealized limit used to define the leading selection rule, the leading common-mode size contribution is projected out.

\section{Doppler links and \texorpdfstring{Michelson-$X$}{Michelson-X} selection rule}

The one-way Doppler observable for a link $s\rightarrow r$ can be written, up to an overall sign convention, as
\begin{equation}
  y_{rs}(t)\equiv \frac{\delta\nu_{rs}(t)}{\nu_0}
  =-\frac{\dd}{\dd t}\delta t_{rs}(t),
  \label{eq:ydef}
\end{equation}
where $r$ labels the receiving spacecraft, $s$ labels the transmitting spacecraft, $\nu_0$ is the nominal laser frequency, and $\delta t_{rs}$ is the optical-time perturbation of the received link.  In natural units the arm length $L$ and the one-way light-travel time are the same quantity, so all one-way propagation delays are written directly as $L$.  We take $\hat{\bm n}_{rs}$ to point from the transmitter $s$ to the receiver $r$; reversing this convention changes only the overall sign of the link response.  For the CM endpoint motion,
\begin{equation}
  \delta t_{rs}^{\rm CM}(t)
  =
  \hat{\bm n}_{rs}\cdot
  \left[
  \delta\bm X_r^{\rm CM}(t)
  -
  \delta\bm X_s^{\rm CM}(t-L)
  \right],
  \label{eq:linkcm}
\end{equation}
Substituting Eq.~\eqref{eq:cmx} directly into the link response gives
\begin{align}
  y_{rs}^{\rm CM}(t)
  =&-\frac{\alpha_A\phi_0}{\omega_\phi}
  \left(\hat{\bm n}_{rs}\cdot\bm{k}\right)
  \Big[
  \cos\!\left(\omega_\phi t-\bm{k}\cdot\bm{x}_r+\varphi\right)
  \nonumber\\
  &\hspace{4.0em}
  -\cos\!\left(\omega_\phi(t-L)-\bm{k}\cdot\bm{x}_s+\varphi\right)
  \Big],
  \label{eq:ylinkcm}
\end{align}
up to the same overall sign convention as Eq.~\eqref{eq:ydef}.  This term is not common mode: it contains both the link-direction projection $\hat{\bm n}_{rs}\cdot\bm{k}$ and the scalar phase sampled at two separated spacecraft.

The endpoint-size term follows from Eq.~\eqref{eq:linksize} in the same way.  For identical endpoints with $q_A(t)=R K_R\phi(t,\bm{x}_A)$,
\begin{align}
  y_{rs}^{\rm size}(t)
  =&\,\omega_\phi R K_R\phi_0
  \Big[
  \sin\!\left(\omega_\phi t-\bm{k}\cdot\bm{x}_r+\varphi\right)
  \nonumber\\
  &\hspace{4.0em}
  -\sin\!\left(\omega_\phi(t-L)-\bm{k}\cdot\bm{x}_s+\varphi\right)
  \Big],
  \label{eq:ylinksize}
\end{align}
again up to the global link-sign convention.  The important difference from Eq.~\eqref{eq:ylinkcm} is that the local amplitude is proportional to $\omega_\phi R K_R\phi_0$ and contains no force-gradient factor $\hat{\bm n}_{rs}\cdot\bm{k}$.  The spatial dependence enters through the phase sampled at the endpoints; it is this phase structure, rather than the local size amplitude alone, that determines whether the signal survives a delayed data combination.

With $y_{rs}$ denoting light received at spacecraft $r$ from spacecraft $s$, and with the spacecraft labels and arm directions chosen as in Fig.~\ref{fig:geometry}, the equal-arm first-generation Michelson-$X$ observable centered on spacecraft 1 can be represented as
\begin{align}
X(t)=&\,y_{31}(t)+y_{13}(t-L)\nonumber\\
&+y_{21}(t-2L)+y_{12}(t-3L)\nonumber\\
&-y_{21}(t)-y_{12}(t-L)\nonumber\\
&-y_{31}(t-2L)-y_{13}(t-3L).
\label{eq:tdix}
\end{align}
We use this first-generation equal-arm expression only as an analytic baseline for the selection rule.  A mission-level calculation with time-dependent unequal arms must propagate the same link perturbations through the appropriate second-generation TDI combinations.
If a perturbation contributes the same common function to all links, $y_{rs}^{\rm com}(t)=Y^{\rm com}(t)$, then the two delay polynomials in Eq.~\eqref{eq:tdix} are identical and the contribution lies in the idealized Michelson-$X$ null direction.

This projection has a simple physical interpretation.  The Michelson-$X$ observable compares two delayed optical paths that start and end at the same spacecraft.  If every link experiences the same scalar-induced timing perturbation in the idealized common-link limit, the delayed copies appear with equal and opposite signs.  The perturbation is then projected into the null space of the laser-noise-canceling data combination.  A scalar-size signal can be large locally and still be strongly reduced after this projection.  Appendix~\ref{app:delay} proves this cancellation by substituting a common endpoint displacement into the delay-polynomial form of $X$.  The appendix is used here to justify keeping only the first common-mode-breaking residuals in Eqs.~\eqref{eq:finitek} and \eqref{eq:unequal}.

When the common-mode limit is broken, the size response reappears only at higher order.  For the equal-arm finite-wave-vector case used in the numerical curves, the residual can be written explicitly.  With $D=e^{-i\omega L}$ and spacecraft 1 taken as the Michelson vertex, substitution of Eq.~\eqref{eq:ylinksize} into Eq.~\eqref{eq:tdix} gives, up to an irrelevant overall sign,
\begin{align}
  T_X^{\rm size}(\omega,\hat{\bm k})
  &\equiv \frac{\tilde X_{\rm size}}{\phi_0}
  \nonumber\\
  &=
  \frac{i\omega R K_R}{2}
  (1-D^4)
  \left(
  e^{-i\bm{k}\cdot\bm{x}_3}
  -
  e^{-i\bm{k}\cdot\bm{x}_2}
  \right)
  \nonumber\\
  &=
  2i\,\omega R K_R\,
  e^{-2i\omega L}
  e^{-i\bm{k}\cdot(\bm{x}_2+\bm{x}_3)/2}
  \nonumber\\
  &\quad\times
  \sin(2\omega L)
  \sin\!\left(\frac{kL\mu_{32}}{2}\right),
  \label{eq:finitek_exact}
\end{align}
where $\mu_{32}\equiv\hat{\bm k}\cdot\hat{\bm n}_{32}$ and $\hat{\bm n}_{32}=(\bm{x}_3-\bm{x}_2)/L$.  Eq.~\eqref{eq:finitek_exact} shows both parts of the suppression: the Michelson delay envelope $\sin(2\omega L)$ and the spatial phase difference between spacecraft 2 and 3.  If one factors out the leading small parameters, the same expression may be written as
\begin{equation}
  T_X^{\rm size}
  =
  \frac{R K_R}{L}\left(\omega L\right)(kL)\,G_X ,
  \label{eq:finitek}
\end{equation}
with the explicit transfer factor
\begin{equation}
  G_X=
  2i e^{-2i\omega L}
  e^{-i\bm{k}\cdot(\bm{x}_2+\bm{x}_3)/2}
  \sin(2\omega L)
  \frac{\sin(kL\mu_{32}/2)}{kL}.
  \label{eq:gxdef}
\end{equation}
Thus $G_X$ is not an additional assumption; it is just the remaining phase, angular, and delay-envelope factor after the powers $(\omega L)(kL)$ have been displayed.  In the equal-arm implementation used for the figures, the finite-$k$ leakage is evaluated in power through the angular factor
\begin{align}
  \mathcal{A}_k(kL)
  &\equiv
  \overline{\sin^2\!\left(\frac{kL\mu}{2}\right)}
  \nonumber\\
  &=
  \frac{1}{2}
  \left[
  1-\frac{\sin(kL)}{kL}
  \right],
  \label{eq:ak}
\end{align}
where $\mu=\hat{\bm k}\cdot\hat{\bm n}_{32}$ and the overbar denotes an isotropic average over the scalar propagation direction.  For $kL\ll1$, $\mathcal{A}_k=(kL)^2/12+O[(kL)^4]$, so the amplitude is linear in $kL$ as stated in Eq.~\eqref{eq:finitek}.  Unequal or time-dependent arms give the analogous residual
\begin{equation}
  T_{X,\Delta L}^{\rm size}(\omega)
  =
  \frac{R K_R}{L}\left(\omega L\right)
  \left(\frac{\Delta L}{L}\right)H_X .
  \label{eq:unequal}
\end{equation}
Unlike $G_X$, there is no single universal closed form for $H_X$ without specifying the unequal-arm pattern and the TDI generation.  For a specified delayed combination, however, it is defined unambiguously.  Let ${\cal P}_X(\{D_a\})$ be the delay polynomial multiplying a common size-link perturbation, where $a$ labels the delay operators appearing in the chosen $X$ combination.  For $L_a=L+\delta L_a$,
\begin{equation}
  \delta D_a=-i\omega D\,\delta L_a ,
  \qquad D=e^{-i\omega L}.
\end{equation}
The first unequal-arm leakage is
\begin{equation}
  \tilde X_{\Delta L}^{\rm size}
  =
  \tilde Y_{\rm size}
  \sum_a
  \left.
  \frac{\partial{\cal P}_X}{\partial D_a}
  \right|_{D_a=D}
  \left(-i\omega D\,\delta L_a\right),
  \label{eq:hxfunctional}
\end{equation}
where $\tilde Y_{\rm size}$ is the common size-link term obtained from Eq.~\eqref{eq:ylinksize}.  Eq.~\eqref{eq:unequal} defines $H_X$ by factoring Eq.~\eqref{eq:hxfunctional} by $(RK_R/L)(\omega L)(\Delta L/L)$.  We keep this term as a scaling estimate for mission-level unequal-arm modeling, rather than as an additional fitted contribution in the baseline equal-arm curves.  Thus the size term after the delayed combination contains additional suppression beyond $R/L$.

The symmetric result follows as the limiting case of the preceding residuals.  In the ideal equal-arm and fully symmetric limit, with identical endpoints, $\Delta L\rightarrow0$, and $kL\rightarrow0$, the breathing of the test masses has the form of a common-mode link perturbation and
\begin{equation}
  \tilde X_{\rm size}(\omega)=0
  \qquad
  (\mathrm{equal\ arms,\ symmetric,\ } kL\rightarrow 0).
  \label{eq:sizezero}
\end{equation}
This is the Michelson-$X$ selection rule for the size channel.

Eqs.~\eqref{eq:finitek} and \eqref{eq:unequal} also identify the controlled ways in which the common-mode projection can be broken.  A finite de Broglie wavelength means that different spacecraft sample slightly different scalar phases.  Unequal or time-dependent arms mean that the delayed common-mode samples are no longer paired in the ideal equal-arm way.  Manufacturing tolerances or deliberately different materials can similarly make the endpoint response nonidentical.  These effects are the channels through which a size-breathing signal can re-enter a TDI data combination, but their order in small parameters is fixed by the selection rule.

\section{Projected sensitivities for LISA, Taiji, and TianQin}

We use the equal-arm Michelson-$X$ noise normalization of Ref.~\cite{Yu2023PRD}.  The following formulas are written in the natural-unit convention fixed above, with the ordinary Fourier frequency $f=\omega/(2\pi)$ used to label one-sided noise spectra.  The mission arm lengths and noise amplitudes quoted in Table~\ref{tab:missions} are converted to the corresponding natural-unit quantities before numerical evaluation:
\begin{align}
  N_X(f)=&\,16\sin^2\tau
  \left[(3+\cos2\tau)S_{\rm acc}(f)+S_{\rm oms}(f)\right],
  \nonumber\\
  \tau=&\,2\pi f L,
  \label{eq:nx}
\end{align}
with
\begin{align}
S_{\rm oms}(f)&=s_{\rm oms}^2(2\pi f)^2
\left[1+\left(\frac{2\,{\rm mHz}}{f}\right)^4\right],\\
S_{\rm acc}(f)&=\left(\frac{s_{\rm acc}}{2\pi f}\right)^2
\left[1+\left(\frac{0.4\,{\rm mHz}}{f}\right)^2\right]\nonumber\\
&\quad\times
\left[1+\left(\frac{f}{8\,{\rm mHz}}\right)^4\right].
\end{align}
Here $N_X(f)$ is the one-sided Michelson-$X$ fractional-frequency noise power spectral density, and $\tau=2\pi fL$ is the dimensionless transfer frequency.  The reference frequencies written in mHz are converted to natural units in the numerical evaluation, just as $L$, $s_{\rm oms}$, and $s_{\rm acc}$ are.  The functions $S_{\rm oms}$ and $S_{\rm acc}$ are the optical-metrology and residual-acceleration contributions in the same fractional-frequency normalization.  The parameter $s_{\rm oms}$ is an optical path-length amplitude noise and is converted to fractional-frequency noise by the factor $2\pi f$, whereas $s_{\rm acc}$ is an acceleration amplitude noise and enters after the corresponding frequency-domain double integration.

The equal-arm model used below is intentionally not presented as a final mission forecast.  It is a controlled comparison of mechanisms using the same normalization for all three missions.  This is the same role played by many sensitivity-curve calculations in the space-interferometer literature: they isolate the leading transfer-function and noise scalings before the complications of orbit files, second-generation TDI, and data-analysis choices are introduced \cite{Prince2002,RobsonCornishLiu2019,Yu2023PRD}.  The resulting curves should be interpreted as a baseline for judging whether the size channel is even competitive with the CM response.

For a response power $R_{X,i}$ to coupling $1/\Lambda_i$, the projected sensitivity is
\begin{equation}
  \left(\frac{1}{\Lambda_i}\right)_{\min}
  = \frac{\rho_{\rm th}}{\phi_0}
  \sqrt{\frac{N_X(f)}{2T_{\rm eff}R_{X,i}(f)}} ,
  \label{eq:sens}
\end{equation}
where $\rho_{\rm th}$ is the signal-to-noise threshold and $T_{\rm eff}$ is the effective integration time.  We take $T_{\rm eff}=T_{\rm obs}$ for coherent integration and $T_{\rm eff}\simeq\sqrt{T_{\rm obs}\tau_{\rm coh}}$ once the observation exceeds the field coherence time.
Eq.~\eqref{eq:sens} follows from the narrow-band signal-to-noise estimate
$\rho^2=2T_{\rm eff}(\phi_0/\Lambda_i)^2R_{X,i}/N_X$, evaluated at the scalar frequency $f_\phi$.  The factor $R_{X,i}$ is dimensionless in the same normalization as the Michelson-$X$ fractional-frequency observable, and $N_X$ is the corresponding one-sided noise power spectral density.  When $T_{\rm obs}<\tau_{\rm coh}$, the signal is treated as a coherent tone.  When $T_{\rm obs}>\tau_{\rm coh}$, independent coherence patches are combined semi-coherently through the effective time used in Eq.~\eqref{eq:sens}.

For the CM part we evaluate the scalar Michelson-$X$ transfer function of Ref.~\cite{Yu2023PRD} directly,
\begin{equation}
  R_{X,i}^{\rm CM}(f)=C_{M,i}^2 v_{\rm DM}^2\,\overline{|X/h|^2},
  \label{eq:rcm}
\end{equation}
where $C_{M,i}$ is the dimensionless material coefficient in Eq.~\eqref{eq:cmcr}, $v_{\rm DM}$ is the adopted halo speed, and $h=C_{M,i}\phi_0 v_{\rm DM}/\Lambda_i$ is the effective scalar-wave strain amplitude used in that transfer-function notation.  The overbar denotes the sky and polarization-basis average used for the scalar propagation direction.  As a cross-check, the numerical sky average reproduces the low-frequency expansion of Ref.~\cite{Yu2023PRD} in the regime $v_{\rm DM}<2\pi fL<1$.

For the endpoint-size residual we use the first nonzero finite-$k$ leakage derived in Sec.~IV.  In the same fractional-frequency normalization as $N_X$, the equal-arm response power implemented in the calculation is
\begin{align}
  R_{X,i}^{\rm size}(f)
  =&
  C_{R,i}^2
  \left(2\omega R\right)^2
  \sin^2(2\omega L)\,
  \mathcal{A}_k(kL),
  \nonumber\\
  k=&\,\omega v_{\rm DM},
  \label{eq:rsize}
\end{align}
with $\mathcal{A}_k$ defined in Eq.~\eqref{eq:ak}.  The factor $2\omega R$ is the Doppler conversion of a local endpoint displacement into fractional frequency, $\sin(2\omega L)$ is the equal-arm Michelson delay envelope, and $\mathcal{A}_k$ is the sky-averaged spatial-phase leakage that breaks the common mode.  The baseline coefficients are $|C_{R,e}|=|C_{R,\gamma}|=1$ and $|C_{R,N}|=m_e/m_N$; the sign of $C_{R,i}$ is dropped because Eq.~\eqref{eq:rsize} is a power response.

The total response used in Eq.~\eqref{eq:sens} is
\begin{equation}
  R_{X,i}(f)=R_{X,i}^{\rm CM}(f)+R_{X,i}^{\rm size}(f).
  \label{eq:rtotal}
\end{equation}
The cross term is not included in the baseline curves: after averaging over the scalar propagation direction and the unknown field phase, the CM and endpoint-size pieces have different angular and delay structures in this simplified equal-arm model.  A phase-resolved orbit-level calculation could keep the full complex amplitudes instead, but Eq.~\eqref{eq:rtotal} is the reproducible power-level prescription used for the present mechanism comparison.

For a network of independent instruments or independent TDI data streams, the natural extension is to add signal-to-noise ratios in quadrature,
\begin{equation}
  \left(\frac{1}{\Lambda_i}\right)_{\min}^{\rm net}
  =
  \frac{\rho_{\rm th}}{\phi_0}
  \left[
  2T_{\rm eff}\sum_a n_a\frac{R_{X,i}^{(a)}(f)}{N_X^{(a)}(f)}
  \right]^{-1/2},
  \label{eq:network}
\end{equation}
where $a$ labels the mission and $n_a$ is the number of independent equal-noise TDI data streams included.  We use $n_a=1$ for a conservative one-Michelson-per-mission network and $n_a=2$ as a simple equal-arm $A/E$ forecast, where $A$ and $E$ denote two orthogonal Michelson-like delayed data combinations treated as independent equal-noise streams in this simplified model.  The latter should be viewed as a controlled sensitivity target, not a substitute for an orbit-resolved second-generation TDI-combination analysis.
The word ``network'' is used in this restricted sense throughout: it denotes a quadrature combination of statistically independent data streams under a common scalar-field model, not a full multi-mission likelihood analysis with mission-specific orbits, duty cycles, calibration systematics, or correlated backgrounds.

\begin{table}[t]
\caption{Mission parameters used in the equal-arm baseline calculation.  They are quoted in conventional engineering units for comparison with mission documents; the numerical code converts them to natural units before using Eqs.~\eqref{eq:nx}--\eqref{eq:network}.}
\label{tab:missions}
\begin{ruledtabular}
\begin{tabular}{lccc}
Mission & $L$ [m] & $s_{\rm oms}$ [pm/$\sqrt{\rm Hz}$] & $s_{\rm acc}$ [m s$^{-2}/\sqrt{\rm Hz}$]\\
\hline
LISA & $2.5\times10^9$ & 15 & $3.0\times10^{-15}$\\
Taiji & $3.0\times10^9$ & 8 & $3.0\times10^{-15}$\\
TianQin & $1.7\times10^8$ & 1 & $1.0\times10^{-15}$\\
\end{tabular}
\end{ruledtabular}
\end{table}

\begin{figure*}[t]
\centering
\includegraphics[width=0.98\textwidth]{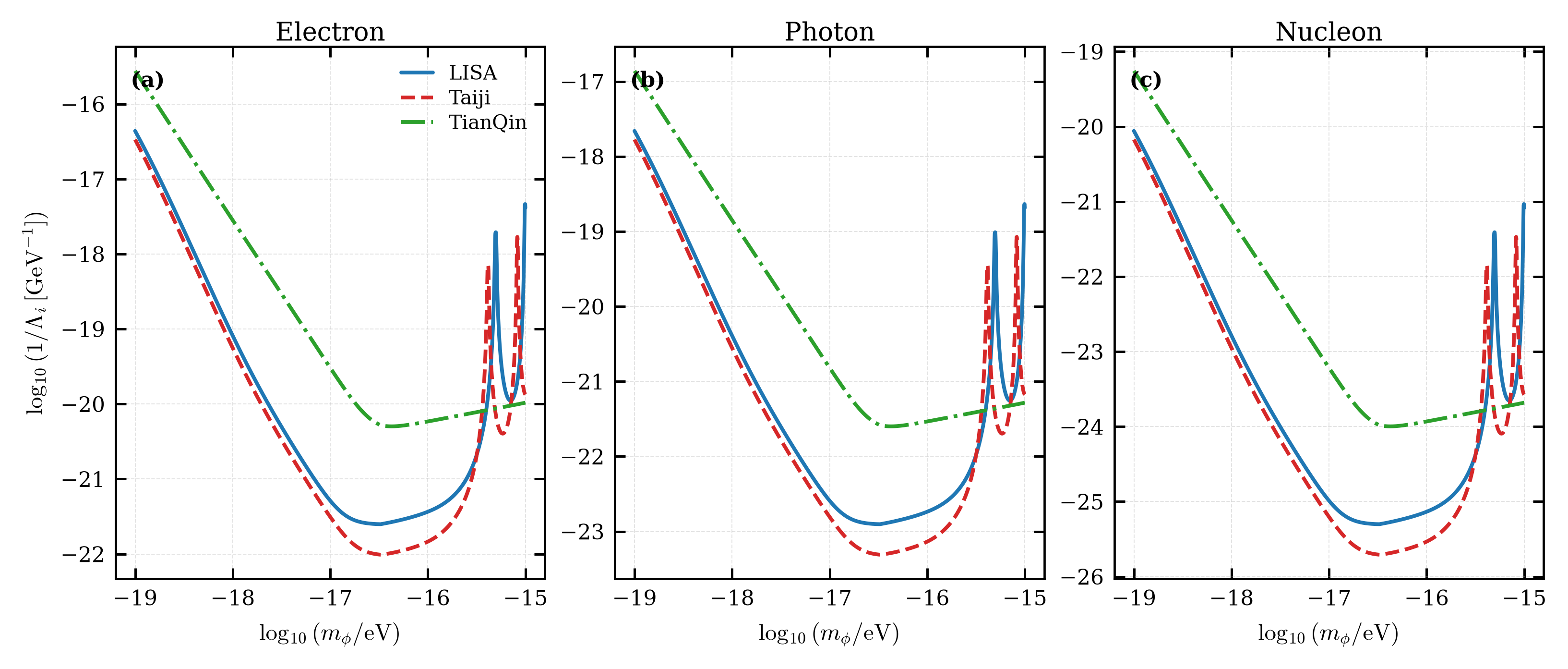}
\caption{Projected total sensitivity for LISA, Taiji, and TianQin in the baseline Au-Pt material model, assuming $\rho_{\rm DM}=0.3\,{\rm GeV\,cm^{-3}}$, $v=10^{-3}$, four years of observation, and unit signal-to-noise threshold.  The total response is the power sum $R_X^{\rm CM}+R_X^{\rm size}$.  In this baseline endpoint model the plotted reach is set mainly by the CM response, so the figure should be read as a mechanism comparison rather than as a final mission data-analysis forecast.}
\label{fig:total}
\end{figure*}

\begin{table*}[t]
\caption{Best single-mission $X$ sensitivity in the range $10^{-19}\le m_\phi/{\rm eV}\le10^{-15}$ for the baseline Au-Pt material model.}
\label{tab:results}
\begin{ruledtabular}
\begin{tabular}{llcc}
Mission & Coupling & $\min\log_{10}(1/\Lambda_i/{\rm GeV}^{-1})$ & $m_\phi^{\rm best}$ [eV]\\
\hline
LISA & $e$ & $-21.602$ & $3.28\times10^{-17}$\\
LISA & $\gamma$ & $-22.903$ & $3.28\times10^{-17}$\\
LISA & $N$ & $-25.301$ & $3.28\times10^{-17}$\\
Taiji & $e$ & $-22.008$ & $3.28\times10^{-17}$\\
Taiji & $\gamma$ & $-23.309$ & $3.28\times10^{-17}$\\
Taiji & $N$ & $-25.707$ & $3.28\times10^{-17}$\\
TianQin & $e$ & $-20.298$ & $4.24\times10^{-17}$\\
TianQin & $\gamma$ & $-21.599$ & $4.24\times10^{-17}$\\
TianQin & $N$ & $-23.997$ & $4.24\times10^{-17}$\\
\end{tabular}
\end{ruledtabular}
\end{table*}

\begin{figure*}[t]
\centering
\includegraphics[width=0.98\textwidth]{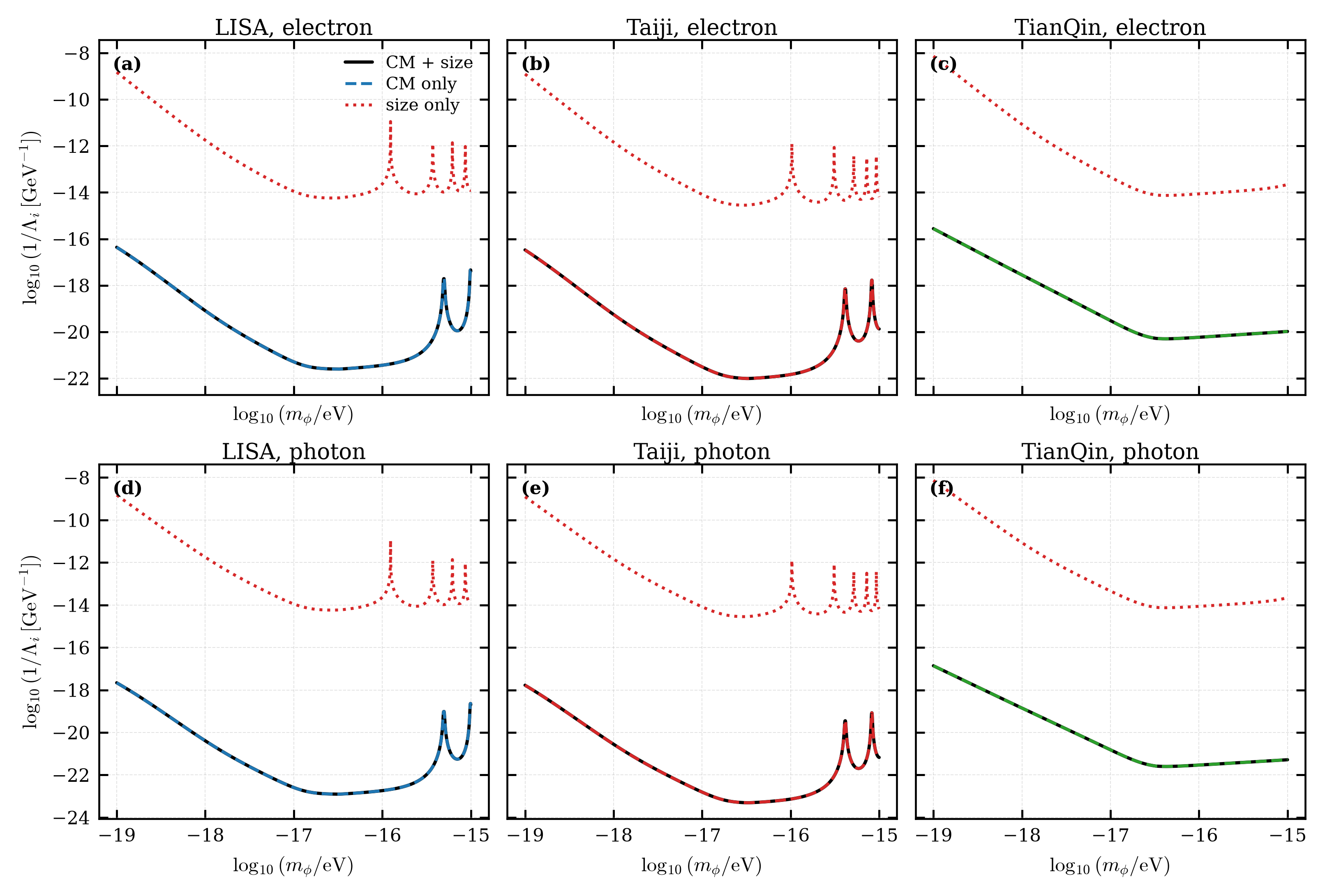}
\caption{Baseline Au-Pt response compared with the counterfactual size-only limit for the electron and photon sectors.  Solid curves include the physical Au-Pt CM response, while dotted curves remove that response by assumption and retain only the first common-mode-breaking endpoint-size residual.  The comparison illustrates how the Michelson-$X$ selection rule affects this particular endpoint model; it is not a complete treatment of optical-bench or beam-splitter-like transduction.}
\label{fig:sizeonly}
\end{figure*}

Table~\ref{tab:results} and Figs.~\ref{fig:total} and \ref{fig:sizeonly} show the same mechanism ordering for all three missions.  Taiji gives the strongest single-mission limits in this equal-arm estimate because its longer arm and smaller optical metrology noise improve the scalar transfer in the relevant band.  TianQin shifts the best point to a somewhat higher mass because of its shorter arm; despite its smaller adopted displacement noises, the shorter baseline makes the CM scalar response weaker at the optimum.  In the endpoint-size model evaluated here, the leading size leakage remains a correction to the CM response rather than a replacement for it.  We do not promote the size-to-CM ratio to a separate exclusion observable, because additional optical-bench or beam-splitter-like paths would require their own instrument transfer functions.

Figure~\ref{fig:sizeonly} should therefore be read diagnostically rather than predictively.  The dotted curves answer a counterfactual question: if the effective scalar charge that drives the CM force were absent, how well could a residual size channel perform?  They do not imply that electron or photon scalar dark matter generically removes the CM channel.  Eq.~\eqref{eq:auptcharge} shows the opposite for Au-Pt: those sectors still carry a material scalar charge, only a smaller one than the nucleon sector.  This distinction is important for comparing with direct searches and equivalence-principle constraints, where the assumed microscopic coupling model controls which material coefficients are present.

\subsection{Network forecast}

The curves in this subsection are generated from the same power-level prescription used for the single-mission forecasts, with statistically independent streams added at the level of signal-to-noise ratio.  For coupling sector $i$ and data stream $c$ of mission $a$, we use
\begin{equation}
  \rho_{a c,i}^2(f)
  =
  2T_{\rm eff}
  \left(\frac{\phi_0}{\Lambda_i}\right)^2
  \frac{R_{a c,i}(f)}{N_{a c}(f)} ,
  \label{eq:stream_snr}
\end{equation}
where $R_{a c,i}$ is the angular-averaged scalar response power for that stream and $N_{a c}$ is its one-sided noise spectrum in the same fractional-frequency normalization.  The network statistic used here is then
\begin{equation}
  \rho_{{\rm net},i}^2(f)
  =
  \sum_a\sum_{c=1}^{n_a}\rho_{a c,i}^2(f).
  \label{eq:network_snr}
\end{equation}
Solving $\rho_{{\rm net},i}=\rho_{\rm th}$ gives Eq.~\eqref{eq:network}.  In the simplified equal-arm $A/E$ forecast we set $R_{a c,i}=R_{X,i}^{(a)}$ and $N_{a c}=N_X^{(a)}$ for two orthogonal equal-noise streams per mission.  Thus the plotted network curves follow from an explicit quadrature sum of independent data streams, not from choosing the best curve by eye.

\begin{figure*}[t]
\centering
\includegraphics[width=0.98\textwidth]{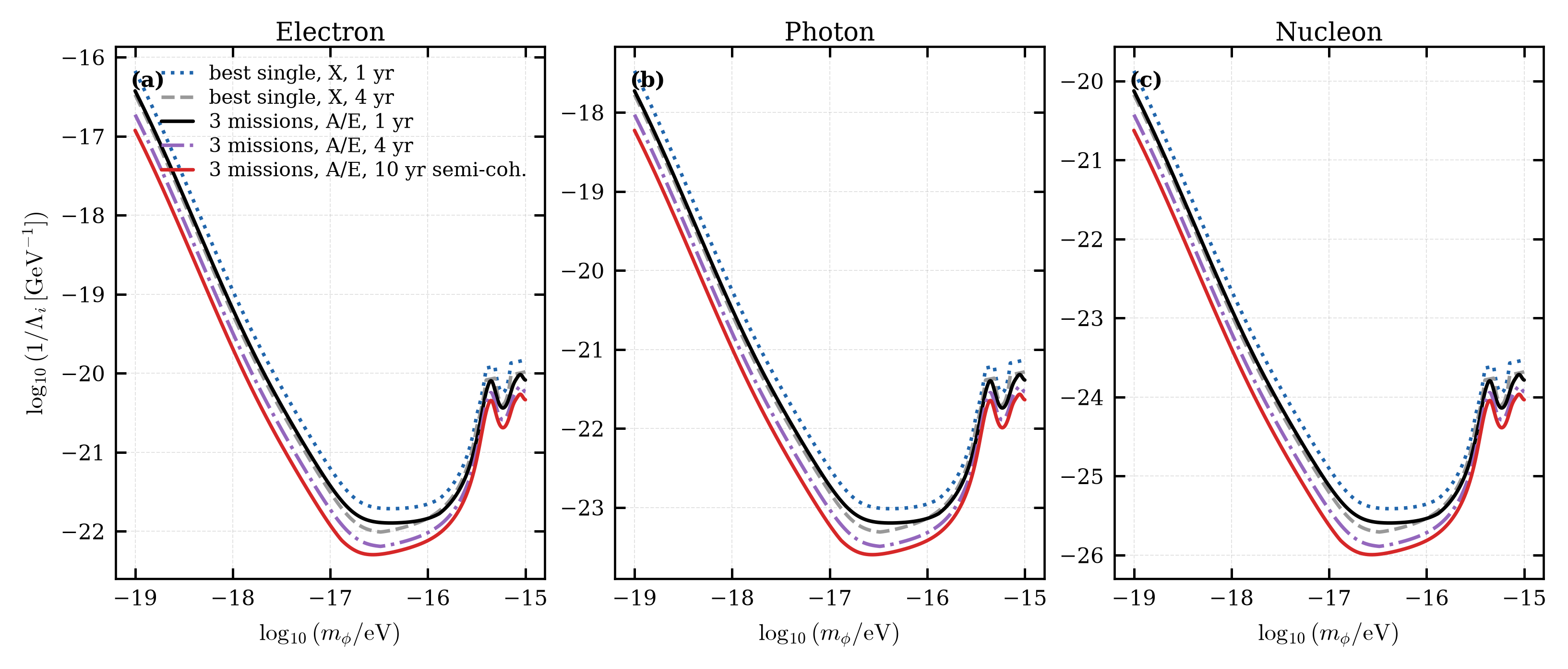}
\caption{Progressively stronger forecast scenarios in the baseline Au-Pt model.  The one-year curves provide a like-for-like comparison with one-year space-detector forecasts in the literature.  Four years is the fiducial mission-duration baseline used for the main sensitivity estimates.  The ten-year curve uses the semi-coherent effective integration time when $T_{\rm obs}>\tau_{\rm coh}$ and is retained only as an extended-duration target, not as the primary claim.}
\label{fig:strongscenarios}
\end{figure*}

The most direct way to strengthen the projected limits without changing the underlying scalar response is to combine independent measurements.  Figure~\ref{fig:strongscenarios} and Table~\ref{tab:strongscenarios} show this bookkeeping effect.  The quoted three-mission curves should therefore be read as a transparent sensitivity projection for independent LISA-, Taiji-, and TianQin-like data streams, not as a claim that a complete operational joint search has already been modeled.  One-year curves are retained because Ref.~\cite{Yu2023PRD} quotes one-year mission forecasts.  Four years is used as the fiducial baseline.  In the mass range where the four-year curves are optimal, $\tau_{\rm coh}$ from Eq.~\eqref{eq:cohtime} is about four years, so the coherent approximation is still a reasonable leading estimate.  Ten years, however, is longer than the coherence time near the optimum; the ten-year row therefore uses the semi-coherent effective time in Eq.~\eqref{eq:sens} and is quoted only as an extended-duration target.  A three-mission network gives a modest gain at the best point in the present equal-arm model, because Taiji already dominates the single-mission optimum.  Including two independent TDI data streams per constellation gives the expected $\sqrt{2}$-level gain, and longer observation gives an additional integration gain only through the coherent or semi-coherent prescription stated above.

\begin{table*}[t]
\caption{Best projected limits for increasingly strong but explicit forecast assumptions.  The last column gives the improvement over the best one-year single-mission $X$ baseline evaluated at the optimum of each scenario.}
\label{tab:strongscenarios}
\begin{ruledtabular}
\begin{tabular}{llccc}
Scenario & Coupling & $\min\log_{10}(1/\Lambda_i/{\rm GeV}^{-1})$ & $m_\phi^{\rm best}$ [eV] & Gain\\
\hline
Best single mission, $X$, 1 yr & $e$ & $-21.713$ & $4.22\times10^{-17}$ & 1.00\\
Three missions, $A/E$, 1 yr & $e$ & $-21.894$ & $4.18\times10^{-17}$ & 1.52\\
Best single mission, $X$, 4 yr & $e$ & $-22.008$ & $3.28\times10^{-17}$ & 2.00\\
Three missions, $A/E$, 4 yr & $e$ & $-22.190$ & $3.28\times10^{-17}$ & 3.04\\
Three missions, $A/E$, 10 yr semi-coh. & $e$ & $-22.295$ & $2.74\times10^{-17}$ & 4.01\\
\hline
Best single mission, $X$, 1 yr & $\gamma$ & $-23.014$ & $4.22\times10^{-17}$ & 1.00\\
Three missions, $A/E$, 1 yr & $\gamma$ & $-23.195$ & $4.18\times10^{-17}$ & 1.52\\
Best single mission, $X$, 4 yr & $\gamma$ & $-23.309$ & $3.28\times10^{-17}$ & 2.00\\
Three missions, $A/E$, 4 yr & $\gamma$ & $-23.491$ & $3.28\times10^{-17}$ & 3.04\\
Three missions, $A/E$, 10 yr semi-coh. & $\gamma$ & $-23.597$ & $2.74\times10^{-17}$ & 4.01\\
\hline
Best single mission, $X$, 1 yr & $N$ & $-25.412$ & $4.22\times10^{-17}$ & 1.00\\
Three missions, $A/E$, 1 yr & $N$ & $-25.593$ & $4.18\times10^{-17}$ & 1.52\\
Best single mission, $X$, 4 yr & $N$ & $-25.707$ & $3.28\times10^{-17}$ & 2.00\\
Three missions, $A/E$, 4 yr & $N$ & $-25.889$ & $3.28\times10^{-17}$ & 3.04\\
Three missions, $A/E$, 10 yr semi-coh. & $N$ & $-25.994$ & $2.74\times10^{-17}$ & 4.01\\
\end{tabular}
\end{ruledtabular}
\end{table*}

The ten-year row is useful as a target for an extended program, but it is not the appropriate baseline for comparison with one-year forecasts.  At the ten-year optimum, $m_\phi\simeq2.7\times10^{-17}\,{\rm eV}$, Eq.~\eqref{eq:cohtime} gives $\tau_{\rm coh}\simeq4.8$ yr for $v=10^{-3}$.  The quoted ten-year gain therefore comes from a semi-coherent combination rather than from assuming a single phase-coherent sinusoid over the full decade.  The main comparison is that a one-year three-mission $A/E$ forecast improves the one-year single-mission baseline by about a factor of $1.5$, while the fiducial four-year three-mission $A/E$ forecast improves it by about a factor of $3$.  The semi-coherent ten-year extension reaches a factor of about $4$ relative to the same one-year baseline.  These gains come from the assumed independent-stream combination, channel multiplicity, and observing time; they do not assume a new optical-bench or beam-splitter transduction model beyond the endpoint response treated here.

\subsection{Comparison with existing limits}

The most direct space-based comparison is Ref.~\cite{Yu2023PRD}.  In the nucleon-dominated dilaton notation used there, $1/\Lambda_N$ corresponds approximately to $\kappa d_g^\ast$, where $d_g^\ast$ is the effective gluonic/nucleon dilaton coefficient used for the single-coupling comparison and $\kappa=\sqrt{4\pi}/M_P=2.9\times10^{-19}\,{\rm GeV}^{-1}$ when the other quark-mass coefficients are set to zero.  Converting our single-mission $X$ forecasts to this notation gives one-year values of order $d_g^\ast\simeq 3.4\times10^{-7}$ for LISA, $1.3\times10^{-7}$ for Taiji, and $7\times10^{-6}$ for TianQin, close to the published LISA/Taiji/TianQin projections in Ref.~\cite{Yu2023PRD}.  The fiducial three-mission, two-channel, four-year scenario gives $1/\Lambda_N=1.3\times10^{-26}\,{\rm GeV}^{-1}$, or $d_g^\ast\simeq4.4\times10^{-8}$, about a factor of $4.7$ below the one-year single-mission Taiji reference point in that normalization.  A ten-year semi-coherent extension would reach $d_g^\ast\simeq3.5\times10^{-8}$, about a factor of six below the same reference point, but that number should be quoted as an extended-duration target rather than as the primary comparison.

\begin{figure*}[t]
\centering
\includegraphics[width=0.98\textwidth]{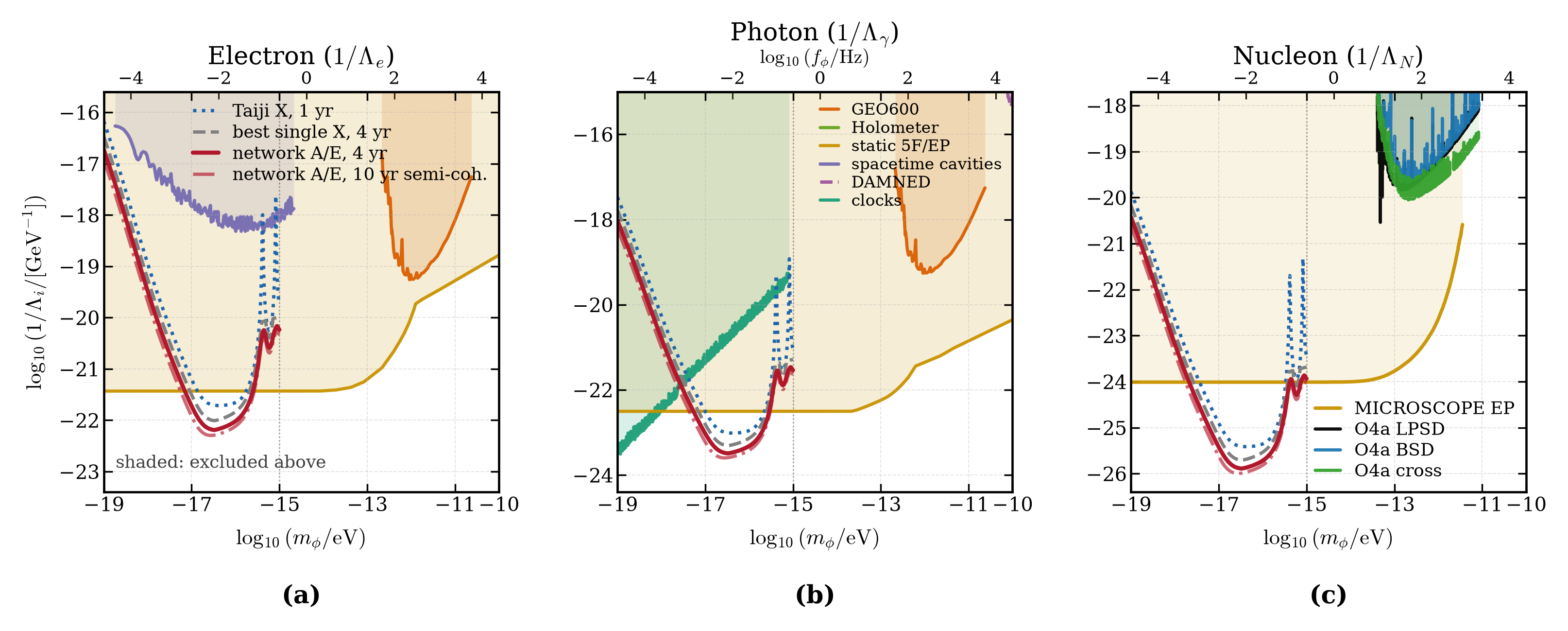}
\caption{External-limit comparison in the manuscript coupling basis, shown over $10^{-19}\le m_\phi/{\rm eV}\le10^{-10}$ to include the low-mass side of ground-based interferometer constraints.  The lower axis gives $m_\phi$ and the upper axis gives the corresponding ordinary oscillation frequency, obtained by restoring SI units as $f_\phi=m_\phi/(2\pi\hbar)$.  Shaded regions denote couplings above the displayed external upper-limit boundary.  The electron and photon panels use the main scalar-coupling axes tabulated in AxionLimits \cite{AxionLimits} and converted with $1/\Lambda_e=\kappa d_{m_e}$ and $1/\Lambda_\gamma=\kappa d_e$ under the single-coupling interpretation of each panel.  Only source curves with tabulated support in this interval are shown; disconnected source domains are not joined, and very short clipped fragments at the edge of the displayed window are omitted.  The cavity-related bounds are plotted source by source: the low-mass space-time-separated cavity result \cite{Filzinger2025PRL} and the higher-mass DAMNED unequal-delay cavity/interferometer result \cite{Savalle2021} should not be read as one continuous experimental contour.  Gold bands in the electron and photon panels show static fifth-force/equivalence-principle projections.  In the nucleon panel, the thin ochre curve shows the MICROSCOPE equivalence-principle result \cite{Touboul2017MICROSCOPE}, including its long-range low-mass plateau, converted from the tabulated $(\lambda,\alpha)$ fifth-force variables with $m_\phi=\hbar c/\lambda$, $g_s^N=\sqrt{\alpha/(1.37\times10^{37})}$, and $1/\Lambda_N\simeq1.07\,g_s^N\,{\rm GeV}^{-1}$.  The O4a curves are from the LVK multi-model search \cite{LVK2025MultiModel}.  These static comparisons are independent of the local dark-matter density and are not statistically equivalent to oscillating-field searches.  Forecast line styles are common to all panels and are listed in the electron panel; these space-detector forecasts are drawn only over $10^{-19}\le m_\phi/{\rm eV}\le10^{-15}$, with the dotted vertical guide marking the upper edge of that forecast window.  The solid red curve is the fiducial four-year three-mission $A/E$ forecast and the dash-dotted red curve is a ten-year semi-coherent extended-duration target.}
\label{fig:external}
\end{figure*}

For the electron and photon sectors we use the main scalar-coupling axes tabulated in the AxionLimits data compilation \cite{AxionLimits}, whose entries reproduce limits from the corresponding clock, spectroscopy, cavity, fifth-force, equivalence-principle, GEO600, and Holometer analyses \cite{VanTilburg2015,Hees2016,Hees2018arXiv,Touboul2017MICROSCOPE,Filzinger2025PRL,Savalle2021,GroteStadnik2019,Vermeulen2021,Aiello2022}.  We convert the dimensionless dilaton variables with $1/\Lambda_e=\kappa d_{m_e}$ and $1/\Lambda_\gamma=\kappa d_e$.  Here $d_{m_e}$ and $d_e$ are the usual dimensionless dilaton couplings to the electron mass and electromagnetic sector, and $\kappa=\sqrt{4\pi}/M_P$ is the inverse reduced Planck scale used in that convention.  This is not the alternative $g_{\phi\gamma}$ axis sometimes shown on scalar-photon plots.  Figure~\ref{fig:external} is shown over $10^{-19}\le m_\phi/{\rm eV}\le10^{-10}$, so that the space-detector forecasts and the lower-mass edge of ground-based interferometer searches can be viewed on the same coupling axes.  The forecast curves themselves are drawn only over the space-detector band, $10^{-19}\le m_\phi/{\rm eV}\le10^{-15}$, corresponding, after restoring SI units for this reporting conversion, to $f_\phi=m_\phi/(2\pi\hbar)\simeq2.4\times10^{-5}$--$0.24\,{\rm Hz}$.  External upper limits are drawn as shaded regions above their converted boundary curves wherever the source data have tabulated support.  Cavity-related limits are kept source-separated: the space-time-separated cavity curve \cite{Filzinger2025PRL} and the DAMNED unequal-delay cavity/interferometer curve \cite{Savalle2021} are not merged into a single experimental contour.

We also show separately labelled static fifth-force and equivalence-principle constraints.  In the electron and photon panels these are the single-coupling lower envelopes tabulated on the scalar-electron and scalar-photon axes.  These static-force constraints are density independent; they are not direct oscillating-field searches and therefore do not share the local-halo-density or coherence assumptions of the clock and cavity curves.  To avoid joining unrelated experimental segments, each ordered source curve is converted first, interpolated only within its own tabulated mass range in $\log m_\phi$--$\log(1/\Lambda)$, and then combined by a pointwise lower envelope.  In the nucleon panel we show the MICROSCOPE equivalence-principle curve as the smooth static reference in the same plotting convention; its low-mass branch is the long-range plateau of the tabulated fifth-force limit \cite{Touboul2017MICROSCOPE}.  The pre-combined ScalarNucleon union file is used only as an audit cross-check, because it switches between limiting experiments and can introduce artificial jumps if plotted as a single contour.  The conversion is given in Appendix~\ref{app:external}.

Ground-based interferometer searches probe a different frequency range.  GEO600, the Holometer, LIGO, and the latest LVK searches are most sensitive at tens of hertz to megahertz frequencies, corresponding roughly to $m_\phi\gtrsim10^{-14}\,{\rm eV}$ for the LIGO band and still higher masses for the Holometer \cite{Vermeulen2021,Aiello2022,Goettel2024PRL,LVK2025MultiModel}.  The right-hand part of Fig.~\ref{fig:external} therefore includes the lower-mass side of these ground-based interferometer constraints, while the space-detector forecast curves are not extrapolated beyond their nominal band.  The LVK O4a multi-model search gives best scalar/dilaton inverse-coupling values around $10^{-21}$--$10^{-20}\,{\rm GeV}^{-1}$ over $m_\phi\simeq4\times10^{-14}$--$8\times10^{-12}\,{\rm eV}$ \cite{LVK2025MultiModel}.  Our fiducial four-year network forecasts reach $1/\Lambda_e\simeq6.5\times10^{-23}\,{\rm GeV}^{-1}$ and $1/\Lambda_\gamma\simeq3.2\times10^{-24}\,{\rm GeV}^{-1}$ near $3\times10^{-17}\,{\rm eV}$.  These numbers should not be interpreted as point-by-point stronger limits, because the mass windows, observables, and coupling conventions differ.  They do show that space interferometers can open a substantially lower-frequency laser-interferometer search window with competitive projected inverse-coupling reach.

\section{Discussion}

The selection rule in Eq.~\eqref{eq:sizezero} clarifies why endpoint-size effects can look important at the single-arm level but become strongly common-mode dominated in a symmetric Michelson-$X$ combination.  It also indicates how to make material-response channels physically relevant.  One route is a model route: arrange for the effective CM scalar charge of the test masses to be negligible, so that the remaining endpoint residual becomes the leading signal.  Another route is an instrumental route: design auxiliary observables that read out size, optical-index, optical-bench, or beam-splitter-like responses directly rather than projecting them into a strongly common-mode-dominated delayed data combination.

The situation is analogous to the ground-based interferometer case.  In GEO600 and LIGO, the signal is not only that a mirror expands.  It is an expansion weighted by the optical geometry, the location of the relevant surface, the symmetry of the arms, and the transfer function from that surface motion to the readout \cite{GroteStadnik2019,Vermeulen2021,Goettel2024PRL}.  In space, the delayed data combinations used for laser-noise suppression add another layer of symmetry filtering.  The present work identifies that filter and estimates the size of the leakage through it.  The main result is therefore a selection rule and a hierarchy of residuals, rather than an unconditional claim of a stronger bound.

\section{Assumptions and limitations}

The present estimates should be regarded as analytic baselines rather than final mission forecasts.  The equal-arm Michelson-$X$ observable is the cleanest setting in which to expose the selection rule, but real constellations use time-dependent arms and ultimately require second-generation TDI.  A publication-level forecast should therefore replace the constant-arm approximation by orbit-resolved LISA, Taiji, and TianQin configurations, propagate the scalar signal through the exact delayed links, and evaluate the response in the same delayed-combination basis used for data analysis.

We also use a single virial velocity scale, $v=10^{-3}$, and quote sky-averaged response powers.  A final search would average over the Galactic velocity distribution and over the stochastic phase of the field.  Near transfer-function zeros, a pointwise sensitivity curve can be visually misleading; frequency-bin averaging or envelope constructions are needed before comparing directly with experimental limits.  These refinements can move detailed curves, especially close to nulls, but they do not remove the symmetry statement that a perfectly common size response is projected out of equal-arm Michelson-$X$.

The material model is deliberately minimal.  We use Au-Pt test-mass scalar charges from the interferometer literature and an adiabatic endpoint-size coefficient.  This is sufficient to test whether a local endpoint-breathing response by itself automatically becomes a leading TDI signal.  It is not a complete model of the spacecraft optical path.  Space interferometers contain optical benches, telescopes, phasemeters, coatings, fiber or free-space routing, and mission-specific beam-combining elements; the exact way in which a scalar-induced beam-splitter-like or optical-index response enters the measured phase depends on the final optical design and calibration scheme.  Those ingredients are important for an instrument paper and should be treated as possible sources of controlled common-mode breaking, not as quantities fixed by the present equal-arm baseline.

Finally, the comparison with external constraints is intentionally limited to conventions that can be matched transparently.  The overlay in Fig.~\ref{fig:external} imports direct oscillating-field electron and photon curves from AxionLimits and converts them to the microscopic inverse scales used here; the same figure includes separately labelled static fifth-force/EP comparisons on the electron and photon axes and the MICROSCOPE scalar-nucleon EP curve, while Appendix~\ref{app:external} records the corresponding conversions.  Existing atomic-clock, cavity, torsion-balance, GEO600, Holometer, and LIGO bounds still use different observables, material coefficients, halo-density assumptions, coherence prescriptions, and statistical thresholds \cite{VanTilburg2015,Savalle2021,GroteStadnik2019,Vermeulen2021,Aiello2022,Fukusumi2023PRD,Goettel2024PRL}.  A final exclusion plot should therefore keep these classes separated rather than drawing a single universal ``best'' bound.  Likewise, a final multi-mission forecast should replace the quadrature network model by mission-specific observing windows, orbits, second-generation TDI data combinations, and correlated-noise checks.  The robust result of this paper is the ordering of mechanisms inside a delayed-link space-detector observable, together with a controlled forecast showing how much independent-stream and observing-time assumptions can improve the baseline.

\section{Conclusions}

We have analyzed scalar dark-matter CM and endpoint-size-breathing responses in a unified one-way-link and TDI-combination treatment for LISA, Taiji, and TianQin.  The main result is a common-mode selection rule: in the equal-arm, fully symmetric, $kL\rightarrow0$ limit, the common endpoint-size contribution is projected into the Michelson-$X$ null direction.  Finite dark-matter wave vector, unequal or time-dependent arms, and material/geometric imperfections reintroduce the endpoint channel through residuals proportional to $(R/L)(\omega L)(kL)$ or $(R/L)(\omega L)(\Delta L/L)$.  Using standard Au-Pt material coefficients, we find that the electron and photon sectors retain nonzero CM scalar charges, so the endpoint-size term does not generically replace the CM response.  Size-only sensitivity curves are therefore useful as counterfactual diagnostics and as guides for future instrument studies, but not as complete projections of all possible optical-path material effects.

For a like-for-like one-year comparison, the single-mission Taiji forecast in this work is close to the Taiji result of Ref.~\cite{Yu2023PRD}; the present calculation is therefore consistent with the existing space-interferometer literature rather than in tension with it.  The stronger limits arise only after the assumptions are made explicit.  Combining LISA, Taiji, and TianQin in two equal-arm $A/E$-like data streams improves the one-year single-mission baseline by about a factor of $1.5$ for one year and by about a factor of $3$ for a fiducial four-year observation, reaching $d_g^\ast\simeq4.4\times10^{-8}$ in the nucleon/dilaton normalization.  A ten-year semi-coherent extension would improve the same baseline by about a factor of $4$, reaching $d_g^\ast\simeq3.5\times10^{-8}$, but this should be treated as an extended-duration target because it exceeds the coherence time near the optimum mass.  The improvement is therefore best understood as a controlled gain from mission network, data-stream multiplicity, and observing time, not as a new dominant size-breathing response.  This separation of baseline CM physics from common-mode-projected size physics provides a clean starting point for orbit-level TDI simulations and future searches for ultralight scalar dark matter with space-based interferometers.

\begin{acknowledgments}
This work is supported by the National Natural Science Foundation of China (Grant No. 12305062) and the Fundamental Research Funds for the Central Universities, Sun Yat-sen University.
\end{acknowledgments}

\section*{Data and code availability}

All numerical results in this manuscript are generated from the accompanying calculation, consistency-check, external-overlay, external-audit, and plotting scripts.  The electron, photon, and nucleon static overlays use the tabulated AxionLimits data compilation \cite{AxionLimits} together with the original experimental references cited in Sec.~V, and the high-frequency nucleon/dilaton comparison uses the public data products released with the LVK O4a multi-model search \cite{LVK2025MultiModel}.  The plotted summary tables and audit tables are written as machine-readable CSV files together with the figures.

\appendix

\section{Delay-operator form of the size-channel projection}
\label{app:delay}

This appendix gives a compact version of the algebra behind Eq.~\eqref{eq:sizezero}.  In natural units the arm length is also the light-travel time.  Let the equal-arm delay operator be
\begin{equation}
  {\cal D}f(t)=f(t-L),
\end{equation}
and let $q_i(t)$ denote the scalar-induced displacement of the relevant optical endpoint on spacecraft $i$ along a given link direction.  For a link $s\rightarrow r$, and up to signs associated with the chosen arm orientation, the size contribution to a one-way fractional-frequency measurement has the schematic form
\begin{equation}
  y_{rs}^{\rm size}(t)
  \simeq {\cal D}\dot q_s(t)-\dot q_r(t).
  \label{eq:app_link}
\end{equation}
This expression is sufficient for the selection rule because it keeps track of the endpoint origin and the light-travel delay.

In the ideal common-mode limit the scalar field is spatially uniform over the constellation and the endpoint responses are identical, so $q_i(t)=q(t)$ for all $i$.  Every directed link then carries the same function,
\begin{equation}
  y_{ij}^{\rm size}(t)=Y(t)
  \equiv \left({\cal D}-1\right)\dot q(t).
\end{equation}
Substituting this into the equal-arm Michelson combination gives
\begin{align}
X_{\rm size}
&=\left(1+{\cal D}+{\cal D}^2+{\cal D}^3\right)Y \nonumber\\
&\quad-\left(1+{\cal D}+{\cal D}^2+{\cal D}^3\right)Y=0 .
\end{align}
The projection is therefore algebraic in this idealized limit.  It does not depend on the amplitude of the local breathing response, only on the fact that the response enters all links in the same delayed way.

The first leakage appears when $q_i$ is not fully common.  For a plane-wave scalar field,
\begin{equation}
  q_i(t)=q_0\cos\left(\omega t-\bm{k}\cdot\bm{x}_i+\varphi\right),
\end{equation}
neighboring spacecraft differ by
\begin{equation}
  q_j-q_i = O(kL)\,q_0 .
\end{equation}
The Doppler measurement supplies one time derivative, giving a factor $\omega q_0$, and the unequal delayed samples in $X$ supply the leading delay-combination factor $1-{\cal D}=O(\omega L)$.  With $q_0\simeq K_R R\phi_0$, we define the leading equivalent-strain amplitude scale as
\begin{equation}
  h_{X,{\rm scale}}^{\rm size}
  =
  K_R\phi_0\frac{R}{L}
  \left(\omega L\right)
  \left(kL\right),
\end{equation}
which is the scaling used in Eq.~\eqref{eq:finitek}.  If the leakage is instead produced by arm-length mismatch, the delay error satisfies $\delta{\cal D}f\simeq-\delta L\,\dot f$, giving the replacement $kL\rightarrow \Delta L/L$ at the same order in $\omega L$.  These estimates explain why the size residual is parametrically smaller than the local one-arm breathing amplitude.

\section{External-limit conversion}
\label{app:external}

This appendix records the normalization used for the external curves in Fig.~\ref{fig:external}.  The scalar-coupling tables imported from AxionLimits are used as a reproducible data compilation of the published clock, cavity, fifth-force, equivalence-principle, GEO600, and Holometer limits cited in Sec.~V \cite{AxionLimits}.  The scalar-electron and scalar-photon tables are tabulated as dimensionless dilaton couplings on the main $d_{m_e}$ and $d_e$ axes.  In the convention of Eq.~\eqref{eq:couplings}, the conversion is $1/\Lambda_e=\kappa d_{m_e}$ and $1/\Lambda_\gamma=\kappa d_e$, with
\begin{equation}
  \kappa=\frac{\sqrt{4\pi}}{M_P}
  =2.90\times10^{-19}\,{\rm GeV}^{-1}.
  \label{eq:app_external_conversion}
\end{equation}
This conversion uses the dimensionless $d_i$ axes, not the alternative $g_{\phi\gamma}$ axis sometimes drawn on scalar-photon summary plots.  For the nucleon comparison we use the single-coupling identification $1/\Lambda_N\simeq \kappa d_g^\ast$, which is the normalization used for the space-detector comparison with Ref.~\cite{Yu2023PRD}.  This identification is approximate because a complete dilaton analysis contains quark-mass and electromagnetic coefficients as well as the gluonic one; here it is used only to place the one-parameter space-interferometer forecasts on a common axis.  The MICROSCOPE scalar-nucleon equivalence-principle source file is tabulated in the fifth-force convention with force range $\lambda$ and relative Yukawa strength $\alpha$ \cite{Touboul2017MICROSCOPE},
\begin{equation}
  V(r)=-\frac{Gm_1m_2}{r}
  \left[1+\alpha e^{-r/\lambda}\right].
  \label{eq:app_yukawa}
\end{equation}
We convert the range to a scalar mass with $m_\phi=\hbar c/\lambda$.  For a scalar nucleon coupling $g_s^N$, AxionLimits uses
\begin{equation}
  \alpha=(g_s^N)^2\,1.37\times10^{37},
  \qquad
  g_s^N=\sqrt{\frac{\alpha}{1.37\times10^{37}}}.
  \label{eq:app_gsn_alpha}
\end{equation}
Following the same AxionLimits normalization used to translate nucleon fifth-force bounds into scalar-coupling axes, we take
\begin{equation}
  d_N=g_s^N\sqrt{1.37\times10^{37}},
  \qquad
  \frac{1}{\Lambda_N}=\kappa d_N .
\end{equation}
Numerically,
\begin{equation}
  \frac{1}{\Lambda_N}
  =1.07\,g_s^N\,{\rm GeV}^{-1}.
  \label{eq:app_nucleon_numeric}
\end{equation}
The pre-combined scalar-nucleon equivalence-principle union file is retained as an audit check of the normalization, but it is not plotted in the main figure because it switches between limiting experiments at high mass and can introduce artificial envelope jumps that are not present in the individual smooth source curves.

The imported electron and photon curves should be read as single-coupling overlays.  Atomic-clock and cavity-related curves are direct oscillating-field searches and retain the halo-density and coherence assumptions of their source analyses.  The main Fig.~\ref{fig:external} shows the part of each converted curve that overlaps $10^{-19}\le m_\phi/{\rm eV}\le10^{-10}$, extending beyond the space-detector forecast window in order to display the low-mass side of ground-based interferometer curves such as GEO600, Holometer, and LVK O4a.  The LISA-, Taiji-, and TianQin-like forecasts are still drawn only over $10^{-19}\le m_\phi/{\rm eV}\le10^{-15}$, where the space-detector noise model and response calculation are used.  The cavity-related entries are plotted source by source rather than as a class envelope: \texttt{Cavities.txt} provides the space-time-separated cavity result \cite{Filzinger2025PRL}, while \texttt{DAMNED.txt} provides the unequal-delay cavity/interferometer result \cite{Savalle2021}.

Fifth-force and equivalence-principle files in the same AxionLimits directories are static-force constraints and do not depend on the local dark-matter density.  The electron and photon gold curves in Fig.~\ref{fig:external} are separately labelled static comparisons: after applying the above coupling conversions, each monotonically ordered source curve is interpolated in $(\log m_\phi,\log(1/\Lambda))$ only over its tabulated mass domain.  When multiple source curves are present on a given axis, the minimum is taken on a common dense mass grid.  Disconnected source domains are kept disconnected when the figure is clipped to the displayed mass range, and very short clipped fragments at the edge of the displayed window are omitted.  For the scalar-nucleon panel, the MICROSCOPE source file is converted directly from $(\lambda,\alpha)$ using Eqs.~\eqref{eq:app_yukawa} and \eqref{eq:app_gsn_alpha}; the long-range plateau tabulated in the same source file supplies the leftward continuation in Fig.~\ref{fig:external}.  Thus the figure compares normalizations and mass coverage, not statistically identical exclusion probabilities.  As a consistency check, the accompanying audit script verifies the AxionLimits column conventions, records header ambiguities in a small number of files, checks that the GEO600 and Holometer single-coupling files are identical in the scalar-electron and scalar-photon directories, records the ScalarNucleon equivalence-principle conversion, and confirms the mass-frequency relation and inverse-coupling columns in the public LVK O4a data products associated with Ref.~\cite{LVK2025MultiModel}.

\bibliographystyle{apsrev4-1}
\bibliography{refs}

@article{GroteStadnik2019,
  author = {Grote, Hartmut and Stadnik, Y. V.},
  title = {Novel signatures of dark matter in laser-interferometric gravitational-wave detectors},
  journal = {Phys. Rev. Research},
  volume = {1},
  pages = {033187},
  year = {2019},
  doi = {10.1103/PhysRevResearch.1.033187},
}

@article{Vermeulen2021,
  author = {Vermeulen, S. M. and others},
  title = {Direct limits for scalar field dark matter from a gravitational-wave detector},
  journal = {Nature},
  volume = {600},
  pages = {424--428},
  year = {2021},
  doi = {10.1038/s41586-021-04031-y},
}

@article{Aiello2022,
  author = {Aiello, L. and Richardson, J. W. and Vermeulen, S. M. and Grote, H. and Hogan, C. and Kwon, O. and Stoughton, C.},
  title = {Constraints on Scalar Field Dark Matter from Colocated Michelson Interferometers},
  journal = {Phys. Rev. Lett.},
  volume = {128},
  pages = {121101},
  year = {2022},
  doi = {10.1103/PhysRevLett.128.121101},
}

@article{Goettel2024PRL,
  author = {G{\"o}ttel, Alexandre S{\'e}bastien and others},
  title = {Searching for Scalar Field Dark Matter with LIGO},
  journal = {Phys. Rev. Lett.},
  volume = {133},
  pages = {101001},
  year = {2024},
  doi = {10.1103/PhysRevLett.133.101001},
}

@article{Michimura2020,
  author = {Michimura, Yuta and Fujita, Tomohiro and Morisaki, Soichiro and Nakatsuka, Hiromasa and Obata, Ippei},
  title = {Ultralight vector dark matter search with auxiliary length channels of gravitational wave detectors},
  journal = {Phys. Rev. D},
  volume = {102},
  pages = {102001},
  year = {2020},
  doi = {10.1103/PhysRevD.102.102001},
}

@article{Yu2023PRD,
  author = {Yu, Jiang-Chuan and Yao, Yue-Hui and Tang, Yong and Wu, Yue-Liang},
  title = {Sensitivity of space-based gravitational-wave interferometers to ultralight bosonic fields and dark matter},
  journal = {Phys. Rev. D},
  volume = {108},
  pages = {083007},
  year = {2023},
  doi = {10.1103/PhysRevD.108.083007},
}

@article{YaoTang2024Stochastic,
  author = {Yao, Yue-Hui and Tang, Yong},
  title = {Probing Stochastic Ultralight Dark Matter with Space-based Gravitational-Wave Interferometers},
  journal = {Phys. Rev. D},
  volume = {110},
  pages = {095015},
  year = {2024},
  doi = {10.1103/PhysRevD.110.095015},
}

@article{YuEtAl2024Gravitational,
  author       = {Yu, Yong and Tang, Yi{-}Dong and Wu, Yu{-}Liang},
  title        = {Detecting Ultralight Dark Matter Gravitationally with Laser Interferometers in Space},
  journal      = {Phys. Rev. D},
  volume       = {110},
  pages        = {023025},
  year         = {2024},
  doi          = {10.1103/PhysRevD.110.023025},
}

@article{LiuEtAl2026EPJC,
  author       = {Liu, Yong-Yong and Zhang, Jing-Rui and Du, Ming-Hui and Liu, He-Shan and Xu, Peng and Zhang, Yun-Long},
  title        = {Detectability of axion-like dark matter for different time-delay interferometry combinations in space-based gravitational wave detectors},
  journal      = {Eur. Phys. J. C},
  volume       = {86},
  number       = {4},
  pages        = {347},
  year         = {2026},
  doi          = {10.1140/epjc/s10052-026-15578-3},
}

@misc{JiangTang2026ULDMInterferometers,
  author = {Jiang, Tingyuan and Tang, Yong},
  title = {Signatures of Ultralight Dark Matter in Space-Based Laser Interferometers},
  year = {2026},
  eprint = {2606.03478},
  archivePrefix = {arXiv},
  primaryClass = {hep-ph}
}

@misc{LVK2025MultiModel,
  author = {{LIGO Scientific Collaboration} and {Virgo Collaboration} and {KAGRA Collaboration} and others},
  title = {Direct multi-model dark-matter search with gravitational-wave interferometers using data from the first part of the fourth LIGO-Virgo-KAGRA observing run},
  year = {2025},
  eprint = {2510.27022},
  archivePrefix = {arXiv},
  primaryClass = {astro-ph.CO}
}

@misc{AxionLimits,
  author = {O'Hare, Ciaran A. J.},
  title = {{AxionLimits}: tabulated limits on axions and ultralight bosonic dark matter},
  year = {2020},
  howpublished = {Zenodo data repository},
  doi = {10.5281/zenodo.3932430},
  note = {Scalar-coupling limit data used for external overlays}
}

@article{Touboul2017MICROSCOPE,
  author = {Touboul, Pierre and others},
  title = {{MICROSCOPE} Mission: First Results of a Space Test of the Equivalence Principle},
  journal = {Phys. Rev. Lett.},
  volume = {119},
  pages = {231101},
  year = {2017},
  doi = {10.1103/PhysRevLett.119.231101}
}

@article{Xu2025Monochromatic,
  author = {Xu, Heng{-}Tao and Yao, Yue{-}Hui and Tang, Yong and Wu, Yue{-}Liang},
  title = {Distinguishing monochromatic signals in {LISA} and {Taiji}: Ultralight dark matter versus gravitational waves},
  journal = {Phys. Rev. D},
  volume = {112},
  pages = {095021},
  year = {2025},
  doi = {10.1103/xlqy-r6n3}
}

@article{GueWolfHees2025LISA,
  author = {Gu{\'e}, Jordan and Wolf, Peter and Hees, Aur{\'e}lien},
  title = {Discriminating scalar ultralight dark matter from quasimonochromatic gravitational waves in {LISA}},
  journal = {Phys. Rev. D},
  volume = {112},
  pages = {115020},
  year = {2025},
  doi = {10.1103/65qp-kvh5}
}

@article{ChenWangLuoShao2026Review,
  author = {Chen, Yuezhe and Wang, Pan-Pan and Wang, Bo and Luo, Rui and Shao, Cheng-Gang},
  title = {The Impact of Dark Matter on Gravitational Wave Detection by Space-Based Interferometers},
  journal = {Universe},
  volume = {12},
  pages = {48},
  year = {2026},
  doi = {10.3390/universe12020048}
}

@misc{YaoEtAl2025Spectral,
  author = {Yao, Yue{-}Hui and Jiang, Tingyuan and Ren, Wenyan and Chen, Di and Tang, Yong and Zhou, Yu{-}Feng},
  title = {Identifying Monochromatic Signals in {LISA} and {Taiji} via Spectral Split: Gravitational Waves versus Ultralight Dark Matter},
  year = {2025},
  eprint = {2508.14655},
  archivePrefix = {arXiv},
  primaryClass = {hep-ph}
}

@article{StadnikFlambaum2015,
  author = {Stadnik, Y. V. and Flambaum, V. V.},
  title = {Searching for Dark Matter and Variation of Fundamental Constants with Laser and Maser Interferometry},
  journal = {Phys. Rev. Lett.},
  volume = {114},
  pages = {161301},
  year = {2015},
  doi = {10.1103/PhysRevLett.114.161301},
}

@article{DereviankoPospelov2014,
  author = {Derevianko, Andrei and Pospelov, Maxim},
  title = {Hunting for topological dark matter with atomic clocks},
  journal = {Nature Physics},
  volume = {10},
  pages = {933--936},
  year = {2014},
  doi = {10.1038/nphys3137},
}

@article{Hees2016,
  author = {Hees, Aur{\'e}lien and Gu{\'e}na, J{\'e}r{\^o}me and Abgrall, M. and Bize, S. and Wolf, P.},
  title = {Searching for an oscillating massive scalar field as a dark matter candidate using atomic hyperfine frequency comparisons},
  journal = {Phys. Rev. Lett.},
  volume = {117},
  pages = {061301},
  year = {2016},
  doi = {10.1103/PhysRevLett.117.061301},
}

@article{DamourDonoghue2010,
  author = {Damour, Thibault and Donoghue, John F.},
  title = {Equivalence principle violations and couplings of a light dilaton},
  journal = {Phys. Rev. D},
  volume = {82},
  pages = {084033},
  year = {2010},
  doi = {10.1103/PhysRevD.82.084033},
}

@article{TintoDhurandhar2014,
  author = {Tinto, Massimo and Dhurandhar, Sanjeev V.},
  title = {Time-delay interferometry},
  journal = {Living Rev. Relativ.},
  volume = {17},
  pages = {6},
  year = {2014},
  doi = {10.12942/lrr-2014-6}
}

@article{Prince2002,
  author = {Prince, Thomas A. and Tinto, Massimo and Larson, Shane L. and Armstrong, J. W.},
  title = {LISA optimal sensitivity},
  journal = {Phys. Rev. D},
  volume = {66},
  pages = {122002},
  year = {2002},
  doi = {10.1103/PhysRevD.66.122002},
}

@article{CornishRubbo2003,
  author = {Cornish, Neil J. and Rubbo, Louis J.},
  title = {LISA response function},
  journal = {Phys. Rev. D},
  volume = {67},
  pages = {022001},
  year = {2003},
  doi = {10.1103/PhysRevD.67.022001},
}

@article{RobsonCornishLiu2019,
  author = {Robson, Travis and Cornish, Neil J. and Liu, Chang},
  title = {The construction and use of LISA sensitivity curves},
  journal = {Class. Quantum Grav.},
  volume = {36},
  pages = {105011},
  year = {2019},
  doi = {10.1088/1361-6382/ab1101},
}

@misc{LISA2017,
  author = {Amaro-Seoane, Pau and others},
  title = {Laser Interferometer Space Antenna},
  year = {2017},
  eprint = {1702.00786},
  archivePrefix = {arXiv},
  primaryClass = {astro-ph.IM}
}

@article{TianQin2016,
  author = {Luo, Jun and others},
  title = {TianQin: a space-borne gravitational wave detector},
  journal = {Class. Quantum Grav.},
  volume = {33},
  pages = {035010},
  year = {2016},
  doi = {10.1088/0264-9381/33/3/035010},
}

@article{Taiji2017,
  author = {Hu, Wen-Rui and Wu, Yue-Liang},
  title = {The Taiji Program in Space for gravitational wave physics and the nature of gravity},
  journal = {Natl. Sci. Rev.},
  volume = {4},
  pages = {685--686},
  year = {2017},
  doi = {10.1093/nsr/nwx116}
}

@article{Marsh2016,
  author = {Marsh, David J. E.},
  title = {Axion cosmology},
  journal = {Phys. Rept.},
  volume = {643},
  pages = {1--79},
  year = {2016},
  doi = {10.1016/j.physrep.2016.06.005},
}

@article{Hui2017Fuzzy,
  author = {Hui, Lam and Ostriker, Jeremiah P. and Tremaine, Scott and Witten, Edward},
  title = {Ultralight scalars as cosmological dark matter},
  journal = {Phys. Rev. D},
  volume = {95},
  pages = {043541},
  year = {2017},
  doi = {10.1103/PhysRevD.95.043541},
}

@article{Arvanitaki2015DM,
  author = {Arvanitaki, Asimina and Huang, JiJi and Van Tilburg, Ken},
  title = {Searching for dilaton dark matter with atomic clocks},
  journal = {Phys. Rev. D},
  volume = {91},
  pages = {015015},
  year = {2015},
  doi = {10.1103/PhysRevD.91.015015},
}

@article{Graham2016DMReview,
  author = {Graham, Peter W. and Irastorza, Igor G. and Lamoreaux, Steven K. and Lindner, Axel and van Bibber, Karl A.},
  title = {Experimental Searches for the Axion and Axion-Like Particles},
  journal = {Ann. Rev. Nucl. Part. Sci.},
  volume = {65},
  pages = {485--514},
  year = {2015},
  doi = {10.1146/annurev-nucl-102014-022120},
}

@article{Bandyopadhyay2025AxionGW,
  author = {Bandyopadhyay, Disha and Borah, Debasish and Das, Nayan and Samanta, Rome},
  title = {High-quality axion dark matter at gravitational wave interferometers},
  journal = {Phys. Rev. D},
  volume = {113},
  pages = {095025},
  year = {2026},
  doi = {10.1103/4mvr-xdc9}
}

@article{TintoArmstrong1999,
  author = {Tinto, Massimo and Armstrong, J. W.},
  title = {Cancellation of laser noise in an unequal-arm interferometer detector of gravitational radiation},
  journal = {Phys. Rev. D},
  volume = {59},
  pages = {102003},
  year = {1999},
  doi = {10.1103/PhysRevD.59.102003}
}

@article{LarsonHellingsHiscock2002,
  author = {Larson, Shane L. and Hellings, Ronald W. and Hiscock, William A.},
  title = {Unequal arm space-borne gravitational wave detectors},
  journal = {Phys. Rev. D},
  volume = {66},
  pages = {062001},
  year = {2002},
  doi = {10.1103/PhysRevD.66.062001}
}

@article{Cutler1998,
  author = {Cutler, Curt},
  title = {Angular resolution of the LISA gravitational wave detector},
  journal = {Phys. Rev. D},
  volume = {57},
  pages = {7089--7102},
  year = {1998},
  doi = {10.1103/PhysRevD.57.7089}
}

@article{VanTilburg2015,
  author = {Van Tilburg, Ken and Leefer, Nathan and Bougas, Lykourgos and Budker, Dmitry},
  title = {Search for Ultralight Scalar Dark Matter with Atomic Spectroscopy},
  journal = {Phys. Rev. Lett.},
  volume = {115},
  pages = {011802},
  year = {2015},
  doi = {10.1103/PhysRevLett.115.011802},
}

@article{Savalle2021,
  author = {Savalle, Etienne and Hees, Aur{\'e}lien and Frank, Florian and Cantin, Etienne and Pottie, Paul{-}Eric and Roberts, Benjamin M. and Cros, Lucie and McAllister, Ben T. and Wolf, Peter},
  title = {Searching for Dark Matter with an Optical Cavity and an Unequal-Delay Interferometer},
  journal = {Phys. Rev. Lett.},
  volume = {126},
  pages = {051301},
  year = {2021},
  doi = {10.1103/PhysRevLett.126.051301},
}

@article{Filzinger2025PRL,
  author = {Filzinger, Melina and Caddell, Ashlee R. and Jani, Dhruv and Steinel, Martin and Giani, Leonardo and Huntemann, Nils and Roberts, Benjamin M.},
  title = {Ultralight Dark Matter Search with Space-Time Separated Atomic Clocks and Cavities},
  journal = {Phys. Rev. Lett.},
  volume = {134},
  pages = {031001},
  year = {2025},
  doi = {10.1103/PhysRevLett.134.031001},
}

@article{Ren2023TDC,
  author       = {Ren, Zhen and others},
  title        = {Taiji data challenge for exploring gravitational wave universe},
  journal      = {Front. Phys.},
  volume       = {18},
  pages        = {64302},
  year         = {2023},
  doi          = {10.1007/s11467-023-1318-y},
}

@article{Fukusumi2023PRD,
  author       = {Fukusumi, Koki and Morisaki, Soichiro and Suyama, Teruaki},
  title        = {Upper limit on scalar field dark matter from LIGO-Virgo third observing run data},
  journal      = {Phys. Rev. D},
  volume       = {108},
  pages        = {095054},
  year         = {2023},
  doi          = {10.1103/PhysRevD.108.095054},
}

@misc{AggarwalHall2022,
  author       = {Hall, Evan and Aggarwal, Nancy},
  title        = {Advanced LIGO, LISA, and Cosmic Explorer as dark matter transducers},
  year         = {2022},
  eprint       = {2210.17487},
  archivePrefix= {arXiv},
  primaryClass = {hep-ph},
  url          = {https://arxiv.org/abs/2210.17487}
}

@article{EstabrookTintoArmstrong2000,
  author       = {Estabrook, F. B. and Tinto, Massimo and Armstrong, J. W.},
  title        = {Time-delay analysis of LISA gravitational wave data: Elimination of spacecraft motion effects},
  journal      = {Phys. Rev. D},
  volume       = {62},
  pages        = {042002},
  year         = {2000},
  doi          = {10.1103/PhysRevD.62.042002},
}

@article{LarsonHiscockHellings2000,
  author       = {Larson, Shane L. and Hiscock, William A. and Hellings, Ronald W.},
  title        = {Sensitivity curves for spaceborne gravitational wave interferometers},
  journal      = {Phys. Rev. D},
  volume       = {62},
  pages        = {062001},
  year         = {2000},
  doi          = {10.1103/PhysRevD.62.062001},
}

@article{Nagano2019PRL,
  author       = {Nagano, Koji and Fujita, Tomohiro and Michimura, Yuta and Obata, Ippei},
  title        = {Axion Dark Matter Search with Interferometric Gravitational Wave Detectors},
  journal      = {Phys. Rev. Lett.},
  volume       = {123},
  pages        = {111301},
  year         = {2019},
  doi          = {10.1103/PhysRevLett.123.111301},
}

@article{Nagano2021PRD,
  author       = {Nagano, Koji and Nakatsuka, Hiromasa and Morisaki, Soichiro and Fujita, Tomohiro and Michimura, Yuta and Obata, Ippei},
  title        = {Axion dark matter search using arm cavity transmitted beams of gravitational wave detectors},
  journal      = {Phys. Rev. D},
  volume       = {104},
  pages        = {062008},
  year         = {2021},
  doi          = {10.1103/PhysRevD.104.062008},
}

@article{LeeNugrohoSpinrath2020EPJC,
  author       = {Lee, Chun-Hao and Nugroho, Chrisna Setyo and Spinrath, Martin},
  title        = {Light dark matter scattering in gravitational wave detectors},
  journal      = {Eur. Phys. J. C},
  volume       = {80},
  number       = {12},
  pages        = {1125},
  year         = {2020},
  doi          = {10.1140/epjc/s10052-020-08692-3},
}

@article{Hees2018arXiv,
  author       = {Hees, Aur{\'e}lien and Minazzoli, Olivier and Savalle, Edwige and Stadnik, Y. V. and Wolf, Peter},
  title        = {Violation of the equivalence principle from light scalar dark matter},
  journal      = {Phys. Rev. D},
  volume       = {98},
  pages        = {064051},
  year         = {2018},
  doi          = {10.1103/PhysRevD.98.064051}
}

@article{Alachkar2025PRL,
  author       = {Alachkar, Ahmad and Fairbairn, Malcolm and Marsh, David J. E.},
  title        = {Dilatonic Couplings and the Relic Abundance of Ultralight Dark Matter},
  journal      = {Phys. Rev. Lett.},
  volume       = {134},
  pages        = {191003},
  year         = {2025},
  doi          = {10.1103/PhysRevLett.134.191003},
}

@article{MillerReview2025,
  author       = {Miller, Andrew L.},
  title        = {Gravitational wave probes of particle dark matter: a review},
  journal      = {Int. J. Mod. Phys. D},
  volume       = {35},
  pages        = {2530005},
  year         = {2026},
  doi          = {10.1142/S0218271825300052},
}

\end{document}